\documentclass{article}

\usepackage{amssymb,amsfonts,amsmath}
\usepackage{cite,enumerate,float,indentfirst}
\usepackage{color}

\def\p{\partial}
\def\tr{{\rm tr}\,}
\def\Tr{{\rm Tr}\,}

\makeatletter
\newdimen\normalarrayskip              
\newdimen\minarrayskip                 
\normalarrayskip\baselineskip
\minarrayskip\jot
\newif\ifold             \oldtrue            \def\new{\oldfalse}
\def\arraymode{\ifold\relax\else\displaystyle\fi} 
\def\eqnumphantom{\phantom{(\theequation)}}     
\def\@arrayskip{\ifold\baselineskip\z@\lineskip\z@
     \else
     \baselineskip\minarrayskip\lineskip2\minarrayskip\fi}
\def\@arrayclassz{\ifcase \@lastchclass \@acolampacol \or
\@ampacol \or \or \or \@addamp \or
   \@acolampacol \or \@firstampfalse \@acol \fi
\edef\@preamble{\@preamble
  \ifcase \@chnum
     \hfil$\relax\arraymode\@sharp$\hfil
     \or $\relax\arraymode\@sharp$\hfil
     \or \hfil$\relax\arraymode\@sharp$\fi}}
\def\@array[#1]#2{\setbox\@arstrutbox=\hbox{\vrule
     height\arraystretch \ht\strutbox
     depth\arraystretch \dp\strutbox
     width\z@}\@mkpream{#2}\edef\@preamble{\halign
\noexpand\@halignto
\bgroup \tabskip\z@ \@arstrut \@preamble \tabskip\z@ \cr}%
\let\@startpbox\@@startpbox \let\@endpbox\@@endpbox
  \if #1t\vtop \else \if#1b\vbox \else \vcenter \fi\fi
  \bgroup \let\par\relax
  \let\@sharp##\let\protect\relax
  \@arrayskip\@preamble}
\def\eqnarray{\stepcounter{equation}%
              \let\@currentlabel=\theequation
              \global\@eqnswtrue
              \global\@eqcnt\z@
              \tabskip\@centering
              \let\\=\@eqncr

 \halign to \displaywidth\bgroup
    \eqnumphantom\@eqnsel\hskip\@centering
    $\displaystyle \tabskip\z@ {##}$%
    \global\@eqcnt\@ne \hskip 2\arraycolsep
         $\displaystyle\arraymode{##}$\hfil
    \global\@eqcnt\tw@ \hskip 2\arraycolsep
         $\displaystyle\tabskip\z@{##}$\hfil
         \tabskip\@centering
    &{##}\tabskip\z@\cr}
\newfont{\hr}{msbm10}
\newfont{\ams}{msam10}

\newdimen\linethick  \linethick=0.4pt
\newdimen\hboxitspace    \hboxitspace=5pt
\newdimen\vboxitspace    \vboxitspace=5pt

\def\fr#1{%
\begin{equation}
\vcenter{
\hrule height\linethick
          \hbox{\vrule width\linethick
                \kern\hboxitspace
                \vbox{\kern\vboxitspace
                      \hbox{$\begin{array}{c}\displaystyle#1
         \end{array}$}%
                      \kern\vboxitspace}%
                \kern\hboxitspace
                \vrule width\linethick}%
          \hrule height\linethick}%
\end{equation}}

\def\beq{\begin{equation}}
\def\eeq{\end{equation}}
\def\ba{\beq\new\begin{array}{c}}
\def\ea{\end{array}\eeq}
\def\be{\ba}
\def\ee{\ea}

\definecolor{red}{rgb}{1,0,0}
\definecolor{orange}{rgb}{1,0.5,0}
\definecolor{violet}{rgb}{0.7,0,1}

\def\theequation{\arabic{section}.\arabic{equation}}

\hoffset= - 1.0in         

\begin{document}

\title{\vspace{.1cm}{\Large {\bf {Orthogonal Polynomials in Mathematical Physics}}\vspace{.2cm}}}
\author{{\bf Chuan-Tsung Chan$^{a}$}\footnote{ctchan@go.thu.edu.tw},
\ {\bf A.Mironov$^{b,c,d,e}$}\footnote{mironov@lpi.ru; mironov@itep.ru},
\ {\bf A.Morozov$^{c,d,e}$}\thanks{morozov@itep.ru},
\ \ and
 \ {\bf A.Sleptsov$^{c,d,e,f}$}\thanks{sleptsov@itep.ru}}
\date{ }

\maketitle

\vspace{-5cm}

\begin{center}
\hfill FIAN/TD-25/17\\
\hfill IITP/TH-21/17\\
\hfill ITEP/TH-35/17
\end{center}

\vspace{2.3cm}

\begin{center}
$^a$ {\small {\it Department of Applied Physics, Tunghai University, Taichung, 40704, Taiwan}}\\
$^b$ {\small {\it Lebedev Physics Institute, Moscow 119991, Russia}}\\
$^c$ {\small {\it ITEP, Moscow 117218, Russia}}\\
$^d$ {\small {\it Institute for Information Transmission Problems, Moscow 127994, Russia}}\\
$^e$ {\small {\it National Research Nuclear University MEPhI, Moscow 115409, Russia }}\\
$^f$ {\small {\it Laboratory of Quantum Topology, Chelyabinsk State University, Chelyabinsk 454001, Russia }}

\end{center}

\vspace{.5cm}

\begin{abstract}
This is a review of ($q$-)hypergeometric orthogonal polynomials and their relation
to representation theory of quantum groups, to matrix models, to integrable theory, and to knot theory.
We discuss both continuous and discrete orthogonal polynomials, and consider their various generalizations.
The review also includes the orthogonal polynomials into a generic framework of ($q$-)hypergeometric functions and their integral representations. In particular, this gives rise to relations with conformal blocks of the Virasoro algebra.
\end{abstract}

\bigskip

\begin{flushright}
{\it To the memory of Ludwig Dmitrievich Faddeev}
\end{flushright}

\bigskip

\tableofcontents

\bigskip

\bigskip

\section{Introduction}

Orthogonal polynomials \cite{BE} are a classical chapter in natural science,
still a somewhat mystical one.
They were originally introduced in XIXth century
in a place, which {may} look strange from today's perspectives:
in the theory of contiued fraction, which is now far away from
the mainstreams in mathematical physics.
Then some of them appeared in study of the hypergeometric series
and entered the textbooks, as the simplest solutions of the most
relevant equations, both in classical and quantum physics
at ordinary energies.
In XXth century, they became for some time the main tool in the study
of matrix models, the basic example and source of inspiration
for modern quantum field and string theories.
Quite recently the hypergeometric polynomials and their
mathematically evident $q$-deformations re-emerged in the
advanced theory of conformal blocks \cite{MMqpain} and
knot invariants \cite{MMSqracah}, they also re-appeared in matrix model in a new avatar \cite{AMMC}. This is one of the immediate
reasons for this paper.
It is basically a review of the well and not-so-well known
facts from the theory of orthogonal polynomials,
with the emphasize on particular aspects needed for a progress
in Painl\'eve and Racah theories, and in matrix models.

We are interested in the interplay of the following subjects:

\begin{itemize}
\item Integral representations (e.g. hypergeometric functions, matrix models).
\item Ward identities (Virasoro/string equations) from invariance under the
change of integration variables.
\item Gauss relations (for hypergeometric functions).
\item 3-term relations.
\item Orthogonal polynomials and matrices.
\item Differential equations (of hypergeometric type, spectral curves).
\item Free-field formalism in representation and quantum field theories (including hypergeometric screening integrals).
\end{itemize}

It is the last item in the list which makes functions of the hypergeometric
type very much distinguished in quantum field theory, while what mattered in classical
physics was rather the second-order differential equations which they often satisfy.
This promotion from classical to quantum level dictates also a change of view
on orthogonal polynomials arising from the hypergeometric series
\be
\sum_k  \frac{\prod_i (a_i)_k}{\prod_j (b_j)_k}\cdot\frac{z^k}{k!}.
\ee
The reduction to polynomials happens whenever any of the parameters $a$ is negative integer,
$a=-n$
since then $(a)_k=\frac{\Gamma(a+k)}{\Gamma(a)}$ vanishes for all $k>n$.
The {resultant from this truncation} is a {polynomial} both in all other parameters $a$ and in the argument $z$.
The difference is only that the dependence on $a$'s is "quantized", while on $z$ is
classical, in fact, $z$ can be considered as an infinitely-large parameter
so that asymptotically $(z)_k \sim z^k$.
Accordingly, a class of polynomials as functions of $z$ is nowadays called "very classical",
while those considered as functions of $a$, just "classical"
(with "quantum" in this context referring to $q$-deformations).

Thus, natural for physical applications are the hypergeometric functions,
and natural for hypergeometric functions is a reduction to polynomials.
However, what is much less natural for these polynomials is orthogonality.
Orthogonal polynomials are distinguished by the peculiar property
known as the 3-term relation, while natural for the hypergeometric polynomials
is to satisfy a $p$-term relation, with $p$ depending on the number of
$a$ and $b$ parameters, i.e. on the type $(r,s)$ of the hypergeometric function ${_r}F_s$.
As we explain in the paper, for the most interesting case of $r=s+1$
(other values of $r-s$ can be achieved by taking certain limits of these reference functions)
the parameter $p$ is equal to 3 for $s\leq 3$, while in general it is rather equal to $2s-3$,
i.e. exceeds $3$ for large $s$.
Moreover, if one insists on polynomiality of the $z$-dependence, i.e. on the
"very classical" polynomials, the $p=3$ bound actually shifts further down to $s=2$,
i.e. to the ordinary hypergeometric functions.
It is {\it this} property that explains at the quantum level the somewhat
mysterious limitations on $s$ allowing appearance of orthogonal polynomials in
the celebrated Askey scheme \cite{Ash,KLS}.
This, in turn, raises a very interesting question of the role of
polynomials with $p$-term relations:
clearly it should exist and be quite important in quantum field theory.

A natural point to look at from this perspective would be matrix-model
applications of orthogonal polynomials, where the natural conjecture could be
that the orthogonality ($p=3$) is related to integrability, the basic property
of matrix model partition functions.
Perhaps, increasing $p$ could be still another natural deformation of integrability.

Another side of the story at $p=3$ {concerns} the orthogonality measure of the polynomials.
Extremely interesting is the case when {the weight function} reduces to a {\it finite} sum
of delta-functions, {and the associated finite set of} orthogonal polynomials actually provide orthogonal matrices.
{Moreover,} when they are hypergeometric, these orthogonal matrices are nothing but
the simplest Racah matrices ($6j$-symbols)!
Thus, the interplay between the hypergeometric functions, $p=3$, and {the discreteness of the weight function}
becomes crucial for the Racah level of representation theory, and, hence, for
the corresponding applications like modular properties of conformal blocks
and knot invariants.

This list of ideas explains the logic of the present paper and its composition.
We begin in sec.2 from reminding the basic properties of orthogonal polynomials
{\it per se}, especially the
continuous-fraction relation between the $3$-term relation and the measure,
which was the birthplace of the theory of orthogonal polynomials in XIXth century.
Then in sec.3, we describe the standard classical viewpoint (Askey scheme)
on the {\it hypergeometric}
orthogonal polynomials, which puts the main emphasize on the second order
differential (or difference, if the $q$-deformation is allowed) equation for them.
In sec.4, we explain the alternative "quantum" viewpoint, where the relations are
generically $p$-term, and only restriction to $s=3$ or $2$ makes $p=3$
and polynomial orthogonal.
In sec.5, we claim that the explanation of $p$-term relations can be achieved
through integral representations of hypergeometric series and is associated
with division of the integration contour in fragments; this, however,
is not very well elaborated in the present paper and deserves further work.
In sec.6, we discuss the discrete orthogonal polynomials, the typical example being the Racah polynomials, and discuss the relation of these latter with the Askey-Wilson polynomials.
In sec.7, we describe various extensions of the classical orthogonal polynomials: to multivariable polynomials, which is essential for constructing the Racah matrices for rectangular representations of $SL_q(N)$, and to Askey-Wilson functions, which solve the same second order differential (difference) equation as the Askey-Wilson polynomials do. We consider the $q$-deformations of the Askey-Wilson functions both in the case of $0<q<1$ and in the case of $|q|=1$. These functions are important for constructing the $6j$-symbols for various infinite-dimensional representations, which we discuss in sect.9 in applications to knot theory.
Finally, secs.8 and 9 remind and develop the standard applications of
orthogonal polynomials respectively to matrix models and to Racah matrices
and knots, in terms which seem most adequate for further thinking about the role of $p=3$.

It is this last application to knots, which brings the present paper
closest to the interests and achievements of Ludwig Faddeev.
Still for us the  main inspiration from Ludwig was the need to search
for deep cross-disciplinary links between different branches of theoretical physics.
Hopefully this paper provides a good example of this kind,
despite there is a very long way from the feelings that we express
to clear and rigorous results.
Making subtle things clear was and remains the trademark of Ludwig Faddeev's
scientific school.
We can only dream about approaching these high standards,
but we try at least to clearly describe the {\it setting}
and the research domain which looks both traditional and extremely intriguing.
We hope that Ludwig would like and appreciate the emerging challenges.

\section{Orthogonal polynomials: a review of standard properties}
\setcounter{equation}{0}

Standard properties of orthogonal polynomials can be found in the books \cite{BE,Chihara}, and those in the $q$-case in \cite{KLS,GaRa} (see also \cite{K13} and references therein).

\subsection{Orthogonal polynomials}
Given a non-decreasing measure function, $\mu(x)$, the orthogonal polynomials can be constructed through  the Gram-Schmidt process,
\be\label{op}
\int P_n(x)\,P_m(x)\,d\mu(x) = ||P_n||^2 \ \delta_{m,n}, \ \ n = 0, 1,2,. \ldots
\ee
where $||P_n||$ is the norm of the $n$-th orthogonal polynomial.
 This set of functions can be viewed as a complete basis of the Hilbert space and there are several equivalent representations:
\begin{itemize}
\item Expression through the moments, $m_n := \int z^n \,d\mu(z)$,
\be
P_n(x) = \det_{(n+1)\times(n+1)}
 \left(\begin{array}{lllll}
x^n & \ldots & m_{n+2} & m_{n+1} & m_n  \\
x^{n-1} & \ldots & m_{n+1} & m_n & m_{n-1} \\
\vdots &   & \vdots & \vdots & \vdots \\
x & \ldots & m_3 & m_2 & m_1\\
1 & \ldots & m_2 & m_1 & m_0
\end{array}\right).
\ee
\item Expression of polynomials through $n$-fold integrals
\be\label{Pmm}
P_n(x) = \dfrac{1}{n!} \int \prod_{i<j}(z_i-z_j)^2 \prod_{i=1}^n (x-z_i) \,d\mu(z_i).
\ee

\item Three-term recursive relation among monic polynomials (i.e. those with unit leading coefficient)
\be\label{3t}
x{\cal P}_n(x) = {\cal P}_{n+1}(x) + K_n {\cal P}_n(x) + R_n {\cal P}_{n-1}(x), \ \ R_n > 0.
\ee
Hereafter, we denote the monic polynomials by calligraphic letters.
\end{itemize}

In fact, if the system of polynomials $P_n(x)$ satisfy the three-term relations,
there is always a measure, for which they are orthogonal and
\be\label{dop}
\sum_{n=0}^\infty \frac{P_n(x) P_n(y)}{||P_n||^2} = \delta(x-y)\cdot\frac{dx}{d\mu(x)}.
\ee
In addition, there exists a continued fraction expression for the generating function of the moments, namely, the resolvent,
\be
d\rho(x) := \left<\frac{dx}{x-z}\right> =
\sum_{k=0}^{\infty} m_k\, \frac{dx}{x^{k+1}} =
 \frac{dx}{x-K_0 - \dfrac{R_1}{x-K_1 -\dfrac{R_2}{x - K_2 -\ldots}}}.
\ee
Hereafter, we denote through $\Big<\ldots\Big>$ the averaging with the measure $\mu(z)$:
\be
\Big<{\cal O}\Big> = \int {\cal O} \ d \mu(z).
\ee

\subsection{Measure from the 3-term relations}

\subsubsection{Uniqueness of measure}

Explicit shape of the measure $d\mu(z)$  depends on the integration contour:
\be
<f> \ = \int_{\cal C} f(z)\ d\mu_{\cal C}(z).
\ee
All measures giving rise to the same average are considered equivalent.
Monic orthogonal polynomials
 are described by a unique orthogonality measure \cite[p.59]{ST} if
\be
\sum_n{1\over\sqrt{R_n}} = \infty
\ee
Furthermore, if the orthogonality measure has a finite support, it is also unique \cite[Ch.2, Th.5.6]{Chihara}.
An example of a non-unique measure can be found in \cite[s.3.2]{K13}.

\subsubsection{Resolvent measure and continued fraction}

If moments of the $z$-variable are known, then the resolvent given by the Stieltjes kind transform,
\be\label{St}
d\rho(x) = {dx} \int_C \frac{d\mu_C(z)}{x-z}=\left<\frac{{dx}}{x-z}\right> = \sum_{k=0}^\infty \frac{<z^k>}{x^{k+1}} {dx},
\ee
can be treated as a measure with the contour surrounding infinity:
\be
<z^k> = - \dfrac{1}{2 \pi i}\oint_\infty x^k\, d\rho(x).
\ee
The inverse transformation from $d\rho(x)$ to $d\mu_C(x)$, including the case of $C=R$
the real line, is more tricky.

If we have a set of monic polynomials, satisfying 3-term relation,
\be
z {\cal P}_n(z) ={\cal P}_{n+1}(z) + K_n  {\cal P}_n(z) + R_n {\cal P}_{n-1}(z),
\ee
then
\be
{\cal P}_0 = 1   \\
{\cal P}_1 = z-K_0   \\
{\cal P}_2 = z^2 - (K_0+K_1)z + (K_0K_1-R_1)   \\
{\cal P}_3 = z^3 - (K_0+K_1+K_2)z^2 + (K_0K_1+K_2K_2+K_1K_2-R_1-R_2)z
- (K_0K_1K_2 - K_2R_1-K_0R_2)   \\
\ldots
\ee
and
\be
<z>\ = K_0,   \\
<z^2>\ = K_0^2 + R_1,   \\
<z^3>\ = K_0^3 + (2K_0+K_1)R_1,   \\
\ldots
\ee
It is easy to check that these moments guarantee orthogonality of
all the polynomials at once,
\be
<{\cal P}_n(z){\cal P}_m(z)> =  \delta_{n,m}\cdot ||{\cal P}_0||^2 \cdot \ \prod_{i=1}^n R_i
\ee
These formulas give rise to the resolvent
in the peculiar form of the continued fraction:
\be
d \rho (x) = \left<\dfrac{dx}{x-z}\right> = \dfrac{dx}{x-K_0 - \dfrac{R_1}{x-K_1
-\dfrac{R_2}{x-K_2 - \dfrac{R_3}{x-K_3 - \ldots}}}} =      \vspace{0.5cm} \\
= \frac{dx}{x-K_0} + \frac{R_1\, dx}{(x-K_0)^2(x-K_1)} +
\left[\frac{R_1^2\, dx}{(x-K_0)^3(x-K_1)^2} + \frac{R_1R_2\,dx}{(x-K_0)^2(x-K_1)^2(x-K_2)}\right]
+ \ldots
\label{dmuinftycf}
\ee

\paragraph{Example.} If all $K_n = 0$ and $R_n = R$ are independent of $n$, then continuous
fraction in (\ref{dmuinftycf}) satisfies $\rho' = 1/(x - R \rho')$ and
\be
d\rho(x) = \frac{x-\sqrt{x^2-4R}}{2R}\,dx
\label{ymeas}
\ee
which is the genus-zero resolvent of the Gaussian matrix model. In fact, this resolvent describes the measure
\be\label{ymeas2}
d\mu(z)=\sqrt{1-z^2/(4R)}\ dz
\ee
at the segment $[-2\sqrt{R},2\sqrt{R}]$ and the system of polynomials orthogonal with respect to this measure is the Chebyshev polynomials of the second type
\be
{\cal P}_n(z)=R^{n/2}\sum_0^{[{n\over 2}]}{(-1)^m(n-m)!\over m!(n-2m)!}\Big({z\over 2\sqrt{R}}\Big)^{n-2m}=P_n^{({1\over 2},{1\over 2})}\Big({z\over 2\sqrt{R}}\Big)
\ee
where we denote $P_n^{(\alpha,\beta)}(z)$ the Jacobi polynomials (see Appendix B).

For orthogonal polynomials, higher resolvents are expressed through the first one
and literally have nothing to do with the matrix model multi-resolvents, their genus expansion
and AMM/EO topological recursion.

\subsubsection{Summing continued fractions with the help of orthogonal polynomials}

However, in general, summation of continued fractions is a far less trivial problem.
In fact, it was the one that originated the interest to orthogonal polynomials. Let us see how this works in the case of the polynomials given on the real axis.
Let us consider the general three-term recurrence relation
\be
\label{3term}
P_{n+1}(z) = \left(A_n z + B_n \right) P_n(z) - C_n P_{n-1}(z), \ n \geq 0.
\ee
A system of orthogonal polynomials $\{P_n(z)\}$ satisfies this relations with the initial conditions
\be
\label{ic1}
P_{-1}(z)=0, \ P_0(z)=1.
\ee
We call these polynomials \textit{primary}.

However one can generate another set of polynomials $\{P^*_n(z)\}$ satisfying the same recurrence relation (\ref{3term}) with different initial conditions
\be
P^*_{n+1}(z) = \left(A_nz+B_n\right)P^*_n(z) - C_n P^*_{n-1}(z), \ n>0  \nonumber \\
P^*_{0}(z)=0, \ \ P^*_1(z)=A_0.
\ee
They are also orthogonal polynomials with respect to the same measure $d\mu(z)$. We call them \textit{secondary} polynomials.
The secondary polynomials are related to the primary ones with the help of Stieltjes transformation
\be
P^*_n(z) = \int_{\mathbb{R}} \dfrac{ P_n(t) - P_n(z) }{t-z} \ d \mu(t).
\ee

In this non-monic case, the resolvent is given by the continued fraction
\be
\label{contfrac}
d \rho(x) = \dfrac{A_0}{A_0x + B_0 - \dfrac{C_1}{A_1x + B_1 - \dfrac{C_2}{A_2x + B_2 - \dots} } }dx
\ee
If this fraction converges, then
\be
\label{limpol}
d\rho(z) = \lim_{n\to \infty} \dfrac{P^*_n(z)}{P_n(z)}\ dz
\ee
and the measure $d\mu(z)$ is related with the continued fraction (resolvent) $d\rho(z)$
by the Stieltjes transformation (\ref{St})
\be\label{meas}
d\rho(x) = dx\int_{\mathbb{R}} \dfrac{ d\mu_{\mathbb{R}}(z) }{x-z}
\ee

In order to find the measure $d\mu(z)$, one should use the inverse Stieltjes transformation due to the Stieltjes-Perron formula:
\be
\boxed{
d\mu(z) = \lim_{\varepsilon\to 0^{+}} \dfrac{ d\rho(z- i \varepsilon) - d\rho(z+ i\varepsilon)}{2\pi i}
}
\ee
One can see that, indeed, the measure (\ref{ymeas2}) can be obtained from the resolvent (\ref{ymeas}) this way.

\paragraph{Example: Hermite polynomials.}

Let us consider the Hermite polynomials and their recurrence relation:
\be
H_{n+1}(z) =  2zH_n(z) - 2n H_{n-1}(z) \nonumber \\
A_n = 2, \ B_n=0, \ C_n=2n.
\ee
According to (\ref{contfrac}), one can write the corresponding continued fraction:
\be
\label{frHerm}
{d\rho(z)\over dz} = \dfrac{2}{2z - \dfrac{2}{2z - \dfrac{4}{2z - \dfrac{6}{2z - \dfrac{8}{2z-\dots}}} } }  \stackrel{z=1/\xi}{=}  \xi+\dfrac{1}{2}\,{\xi}^{3}+\dfrac{3}{4}\,{\xi}^{5}+{\frac {15}{8}}{\xi}^{7}+{\frac {105}{16}}{\xi}^{9}+{\frac {945}{32}}{\xi}^{11}+O \left( {\xi}^{12} \right)
\ee
On the other hand, the first few Hermite polynomials and the corresponding secondary polynomials are explicitly given as,
\be
\begin{array}{lcl}
P_0(z) = 1, && P^{*}_0(z) = 0, \\
P_1(z) = 2z, && P^{*}_1(z) = 2, \\
P_2(z) = 4z^2-2,  &&  P^{*}_2(z) = 4z, \\
P_3(z) = 8z^3-12z,  &&  P^{*}_3(z) = 8z^2-8, \\
P_4(z) = 16z^4-48z^2+12,  &&  P^{*}_4(z) = 16z^3-40z, \\
P_5(z) = 32z^5-160z^3+120z.  &&  P^{*}_5(z) = 32z^4-144z^2+64. \\
\end{array}
\ee
By computing the first few ratios between the secondary and primary polynomials,
\[ \begin{array}{rcl}
\dfrac{P^{*}_1(z)}{P_1(z)} = \dfrac{2}{2z} &\stackrel{z=1/\xi}{=}& {\bf \xi } \nonumber \\
\dfrac{P^{*}_2(z)}{P_2(z)} = \dfrac{4z}{4z^2-2} = \dfrac{2}{2z-\dfrac{2}{2z}}
&\stackrel{z=1/\xi}{=}& {\bf \xi+\dfrac{1}{2}\xi^3}+\dfrac{1}{4}\xi^5+O(\xi^7) \nonumber \\
\dfrac{P^{*}_3(z)}{P_3(z)} = \dfrac{8z^2-8}{8z^3-12z} = \dfrac{2}{2z-\dfrac{2}{2z-\dfrac{4}{2z}}}   &\stackrel{z=1/\xi}{=}& {\bf \xi+\dfrac{1}{2}\xi^3+\dfrac{3}{4}\xi^5}+\dfrac{9}{8}\xi^7+\dfrac{27}{16}\xi^9+O(\xi^{11}) \nonumber \\
\dfrac{P^{*}_4(z)}{P_4(z)} = \dfrac{16z^3-40z}{16z^4-48z^2+12} = \dfrac{2}{2z-\dfrac{2}{2z-\dfrac{4}{2z-\dfrac{6}{2z}}}}
&\stackrel{z=1/\xi}{=}& {\bf \xi+\dfrac{1}{2}\xi^3 +\dfrac{3}{4}\xi^5 +\dfrac{15}{8}\xi^7}+\dfrac{81}{16}\xi^9+O(\xi^{11})
\end{array}\]
one can see that the first $n$ terms of $\dfrac{P^{*}_n(z)}{P_n(z)}$ coincide with the first $n$ terms of the series (\ref{frHerm}), which illustrates (\ref{limpol}).
It is also clear that, in the limit of $n\to\infty$, this ratio approaches to the following series,
\be
\dfrac{d\rho(z)}{dz} = \dfrac{P^{*}_n(z)}{P_n(z)} \ \to {1\over\pi}\sum_k \Gamma\Big(k+{1\over 2}\Big)z^{-2k-1}.
\ee
The Borel sum of this series is
\be
{1\over\pi}\int_0^\infty dt \ e^{-t} \ \sum_k \Gamma\Big(k+{1\over 2}\Big)z^{-2k-1} \ {t^k\over k!}={1\over\pi}\int_0^\infty{dt e^{-t}\over\sqrt{z^2-t}}.
\ee
This integral converges at pure imaginary $z=i\zeta$ and is equal (upon substitution $t\to y^2+2y\zeta$) to
\be
{d\rho(z)\over dz} = {1\over i\pi} \ e^{\zeta^2} \ \hbox{erf}(-\zeta)={1\over i\pi} \ e^{-z^2} \ \hbox{erf}(iz)={1\over i\pi}\int_{-\infty}^{\infty}{e^{-x^2}\over z-x} \ dx
\ee
i.e. from (\ref{meas}), the measure is the Gaussian one, $d\mu_{\mathbb{R}}(x)=e^{-x^2}$.

\subsection{Orthogonal polynomials of discrete variable}

If the measure $\mu(x)$ in (\ref{op}) is a non-decreasing step function with jumps
$\mu_j > 0$ at $x=x_j$, $j=1,2, \ldots, N$,
\be\label{discmeas}
d\mu(x)   = \sum_{j=1}^N \mu_j \ \delta(x-x_j) \ dx.
\ee
then (\ref{op}) acquires the form
\be\label{dopr}
\sum_{j=1}^N P_n(x_j) \ P_m(x_j)\ \mu_j = ||P_n||^2 \ \delta_{m,n}
\ee
The counterpart of (\ref{dop}) in this case is
\be\label{dual}
\sum_n {P_n(x_i) P_n(x_j) \over ||P_n||^2} = {\delta_{i,j}\over \mu_j}.
\ee
The usual boundary conditions
\be
P_{-1}(x_j)=0,\ \ \ \ P_0(x_j)=1,\ \ \ \ \forall j
\ee
is often added with
\be
P_N(x_j)=0,\ \ \ \ \ \forall j,
\ee
at some $N$. An example of a system of such polynomials can be found in s.\ref{Racah}.

In the case of polynomials orthogonal with discrete measure (\ref{dopr}),
\be
U_{ij} = P_i(x_j)\cdot \frac{\sqrt{\mu_j}}{||P_i||},
\ee
is an $N\times N$ orthogonal (unitary) matrix.

\section{Classical orthogonal polynomials}
\setcounter{equation}{0}
\subsection{Generalities}

Considering orthogonal polynomials with the measure of generic type is too unspecified problem, which, hence, allows one to make only very general statements about the polynomials, which we listed in the previous section. There are, however, various possibilities of specifying the family of orthogonal polynomials. In this section, we briefly describe the family that is called the classical orthogonal polynomials. Following T. Koornwinder, we differentiate between the {\it classical} and {\it very classical} polynomials: the latter ones are those considered in the standard text-books (see Appendix B), while the classical polynomials are those described by the Askey scheme (see s.\ref{As}).
In fact, all classical polynomials can be associated either with the hypergeometric functions or with their $q$-deformation. Hence, sometimes we call polynomials of such a type hypergeometric.

Being hypergeometric, all classical polynomials have integral representations which belong to two large (intersecting) classes:
\begin{itemize}
\item The very classical polynomials are all specializations of the Jacobi polynomials, which are standard hypergeometric functions $_2F_1(x)$
and, hence, can be realized as one-dimensional integrals,
\be
P_n(x) = \int (x-z)^n d\nu_n(z),
\ee
with some measure $d\nu_n(z)$, which depends on $n$ in a relatively simple way.
\item  Other classical polynomials are of the Askey-Wilson (AW) type (hence, the name AW type polynomials), where the dependence on $x$ and $n$ is in distinct factors (and can be recast to the powers of linear functions in $z$ as we explain below):
\be
P_n(x) = \int z^n \psi(z,x) d\eta_n(z).
\ee
\end{itemize}
These two families of polynomials are related with the idea that, in hypergeometric case, the polynomial is of the form like
\be
\sum_{k=0}^n \frac{n!}{k!(n-k)!} \frac{\Gamma(u+k)}{\Gamma(u)}\cdot s_k z^k,
\ee
i.e. is a polynomial both in $z$ (the very classical polynomial) and in $u$ (the remaining classical polynomials),
but the methods to make them orthogonal differ in the two classes.

\subsection{Classical orthogonal polynomials as hypergeometric polynomials}

The hypergeometric functions of one variable are defined as (see Appendix A)
\be
\phantom._rF_s\left(\begin{array}{c} a_1,\ldots, a_r \\ b_1,\ldots,b_s \end{array}\Big|\,z\right)
:= \sum_{k=0}^\infty \frac{(a_1)_k\ldots (a_r)_k}{(b_1)_k\ldots (b_s)_k}\cdot\frac{z^k}{k!}.
\ee
These functions become polynomials when any of $a_i = -n$:
in these cases, they are polynomials both in $z$ (the very classical case)
and in any other variable $a_j$ (the AW type case).

\subsubsection{The very classical case}

The hypergeometric functions satisfy the following equations
\be
\left(z\frac{\p }{\p z} + a_i\right)\phantom._rF_s\left(\begin{array}{c} a_1,\ldots,a_i,\ldots, a_r \\ b_1,\ldots,b_s \end{array}\Big|\,z\right) = a_i
\phantom._rF_s\left(\begin{array}{c} a_1,\ldots,a_i+1,\ldots, a_r \\ b_1,\ldots,b_s \end{array}\Big|\,z\right),   \\
\left(z\frac{\p}{\p z} + b_j-1\right) \phantom._rF_s\left(\begin{array}{c} a_1,\ldots, a_r \\ b_1,\ldots,b_j,\ldots,b_s \end{array}\Big|\,z\right) = (b_j-1)
\phantom._rF_s\left(\begin{array}{c} a_1,\ldots, a_r \\ b_1,\ldots,b_j-1,\ldots,b_s \end{array}\Big|\,z\right),
\label{dilats}
\ee
and
\be
\frac{\p}{\p z} \phantom._rF_s\left(\begin{array}{c} a_1,\ldots, a_r \\ b_1,\ldots,b_s \end{array}\Big|\,z\right) = \frac{\prod a_i}{\prod b_j}
\phantom._rF_s\left(\begin{array}{c} a_1+1,\ldots, a_r+1 \\ b_1+1,\ldots,b_s+1 \end{array}\Big|\,z\right).
\label{streq}
\ee
From the point of view of matrix models these are just the lowest of
the Virasoro constraints, see \cite{MMqpain}.

From these equations, we deduce the following consequences:
\begin{itemize}
\item the hypergeometric differential equations of order ${\rm max}(r,s+1)$,
which follows from multiplying (\ref{dilats}) for $a_i$ shifts and comparing with
the product of (\ref{streq}) and all (\ref{dilats}) for $b_j$;
\item linear algebraic relations (Gauss' contiguous relations) between $F$ with shifted parameters $a$ and $b$.
\end{itemize}
The Gauss relations imply 3-term relations for the case $(r,s)=(2,1)$ which can be recast into the form (\ref{3t}) provided $a_1=-n$ and $a_2=n+\alpha$, where $\alpha$ is an arbitrary constant. This means that the polynomials $\phantom._2F_1\left(\begin{array}{c} -n,n+\alpha \\ \beta \end{array}\Big|\,z\right)$ satisfy 3-term relations and, hence, they are
orthogonal polynomials. These are the most general very classical orthogonal polynomials, the Jacobi polynomials (see Appendix B), which we denote as $P_n^{(a,b)}(z)$:
\be
P_n^{(a,b)}(x) := \dfrac{(a+1)_n}{n!} \ {_2}F_1\left( \begin{array}{c} -n, \ \ a+b+n+1  \\  a+1  \end{array} \Big|\ \frac{1}{2}(1-x) \right).
\ee

\subsubsection{Askey-Wilson polynomials\label{AW}}

Of all the AW type polynomials, the most general are the AW polynomials themselves \cite{AW}, all other are obtained upon suitable specializations. Hence, we describe here the AW polynomials and their properties following \cite[sec.7.5]{GaRa}. In particular, for the sake of generality, we discuss the $q$-deformation of the polynomials, which is not too much different from the AW polynomials at $q=1$. All the definitions of the basic hypergeometric series, $_r\phi_s$, and other $q$-functions can be found in Appendix A.

The AW polynomials are defined through the balanced ${_4\phi_3}$ basic hypergeometric series,
\be\label{AWpol}
P_n^{AW}(x=\cos\theta|a,b,c,d) := a^{-n} (ab,ac,ad;q)_n\cdot\,{_4\phi_3}\left(\begin{array}{c}
q^{-n},abcdq^{n-1},ae^{i\theta},ae^{-i\theta}\\ ab,ac,ad\end{array}\Big|z=q\right).
\ee
They are symmetric in $a,b,c,d$, which can be proved using the Sears transformation formula between two terminating balanced $_4\phi_3$ series \cite{GaRa}
\be
\phantom{.}_4\phi_3\left(\begin{array}{c}
q^{-n},a,b,c\\ d,e,f\end{array}\Big|z=q\right)=a^n\ {(e/a,f/a;q)_n\over (e,f;q)_n}
\phantom{.}_4\phi_3\left(\begin{array}{c}
q^{-n},a,d/b,d/c\\ d,q^{1-n}a/e,q^{1-n}a/f\end{array}\Big|z=q\right)
\ee
The polynomials satisfy the duality relation
\be\label{duality}
P_n^{AW}(aq^m|a,b,c,d)=P_m^{AW}(\hat a q^n|\hat a,\hat b,\hat c,\hat d),\ \ \ \ \ \hat a:=\sqrt{abcd\over q},\ \hat b:={ab\over \hat a},\ \hat c:={ac\over\hat a},\ \hat d:={ad\over\hat a}
\ee
The 3-term relation is of the form
\be
2xP^{AW}_n(x) = A_nP^{AW}_{n+1}(x)+B_nP^{AW}_n(x)+C_nP^{AW}_{n-1}(x),
\ee
with $P^{AW}_{-1}(x)=0$, $P^{AW}_0(x)=1$, and
\be
A_n = \frac{1-abcdq^{n-1}}{(1-abcdq^{2n-1})(1-abcdq^{2n})},   \\
C_n = \frac{(1-q^n)(1-abq^{n-1})(1-acq^{n-1})(1-adq^{n-1})}
{(1-abcdq^{2n-2})(1-abcdq^{2n-1})}\cdot(1-bcq^{n-1})(1-bdq^{n-1})(1-cdq^{n-1}),  \\
B_n = a+\frac{1}{a} -A_n\frac{(1-abq^n)(1-acq^n)(1-adq^n)}{a}
- C_n\frac{a}{(1-abq^{n-1})(1-acq^{n-1})(1-adq^{n-1})}.
\ee
The orthogonality condition at $|q|<1$ and with none of the pairwise products of the parameters $a$,$b$, $c$, $d$ equal to $q^{-k}$, $k\in\mathbb{Z}_{\ge 0}$ is
\be\label{orAWg}
{1\over 2\pi i}\int_{\cal C} P^{AW}_m\Big({z+z^{-1}\over 2}\Big)P^{AW}_n\Big({z+z^{-1}\over 2}\Big){(z^2,z^{-2};q)_\infty\over
(az,az^{-1},bz,bz^{-1},cz,cz^{-1},dz,dz^{-1};q)_\infty}{dz\over z}= 2\frac{\delta_{m,n}}{h_n}\int_{-1}^1 w(x)dx,
\ee
where ${\cal C}$ is the unit circle traversed in positive direction with suitable deformations to separate the sequences of poles converging to zero from the sequences of poles diverging to $\infty$ and
\be\label{AWmu}
w(x|a,b,c,d) = \frac{h(x|1,-1,\sqrt{q},-\sqrt{q})}{h(x|a,b,c,d)\sqrt{1-x^2}},
\ee
\be\label{h}
h(x|a_1,\ldots,a_m) := \prod_{i=1}^m  (a_ie^{i\theta},a_ie^{-i\theta};q)_\infty = \prod_{i=1}^m \prod_{n=0}^{\infty}(1-2a_ixq^n+a_i^2q^{2n}),
\ee
\be
\int_{-1}^1 w(x)dx = \frac{2\pi(abcd;q)_\infty}{(q,ab,ac,ad,bc,bd,cd;q)_\infty},
\ee
\be
h_n =
{(1-abcdq^{2n-1})\over (1-abcdq^{-1})}{(abcdq^{-1};q)_n\over(q,ab,ac,ad,bc,bd,cd;q)_n}.
\ee
If two, one or zero pairs of complex conjugate among the parameters $a$, $b$, $c$, $d$ with remaining parameters real
and $|a|$, $|b|$, $|c|$, $|d|$ are all less than 1, the contour is just the unit circle and the orthogonality relation takes the form
\be
\int_{-1}^1 P^{AW}_m(x)P^{AW}_n(x) w(x)dx= \frac{\delta_{m,n}}{h_n}\int_{-1}^1 w(x)dx,
\label{AWmeasure}
\ee
Assume that one of the parameters, say, $|a|>1$. Then, one can again deform the contour to the unit circle and add residues at the poles at all $aq^k$ lying outside the circle, the result being
\be\label{AWmeasure2}
\int_{-1}^1 P^{AW}_m(x)P^{AW}_n(x) w(x)dx+2\pi\sum_k  P^{AW}_m(x_k)P^{AW}_n(x_k)w_k= \frac{\delta_{m,n}}{h_n}\int_{-1}^1 w(x)dx
\ee
where
\be
w_k={1-a^2q^{2k}\over 1-a^2}{(a^{-2};q)_\infty\over (q,ab,ac,ad,b/a,c/a,d/a;q)_\infty}{(a^2,ab,ac,ad;q)_k\over (q,aq/b,aq/c,aq/d;q)_k}
\Big({q\over abcd}\Big)^k
\ee
and the sum runs over the points $x_k=(aq^k+q^{-k}/a)/2$ with $aq^k$ lying outside the unit circle.

\subsection{Askey scheme\label{As}}

As we already noticed, the Askey-Wilson polynomials are the maximal ones within the so-called ($q$)-Askey scheme \cite{Ash,KLS}, which describes the classical orthogonal polynomials. These polynomials all have a representation through the Rodriguez formula, and their measure is determined from the Pearson operator equation \cite{KLS}.

Let us start with not orthogonal polynomials with $q=1$. Consider polynomials that solve an eigenvalue problem of the form (all classical polynomials are of this kind),
\be\label{ev}
A(x)P_n(Q(x+\gamma))+B(x)P_n(Q(x))+C(x)P_n(Q(x-\gamma))=\lambda_nP_n(Q(x)),
\ee
where $Q(x)$ is a second degree polynomial and $\gamma$ is a parameter. Then, it turns out that there are only 13 families of orthogonal polynomials that solve the eigenvalue problem (\ref{ev}). They depend on at most four parameters and all are obtained by the maximal member of the family, the Askey polynomial. Hence, all of them are expressed through hypergeometric functions, at most, $_4F_3$. An important property of this kind of polynomials is that all they can be expressed from the measure through the Rodriguez formula.

The same scheme is equally well applicable in the $q$-deformed case (note that, apart from limit $q\to 1$, there is an interesting limit $q=-1$, where more ``classical'' polynomials emerge \cite{VZ}). In this case, we consider the eigenvalue problem
\be
\phi(x){\hat H}_q {\hat H}_{1/q} (x)+\psi(x){\hat H}_q P_n(x)=\lambda_nP_n(qx+\omega)
\ee
where ${\hat H}_q$ is the Hanh operator:
\be
{\hat H}_q f(x):={f(qx+\omega)-f(x)\over qx+\omega-x}={\hat Df(x)-f(x)\over \hat Dx-x}
\ee
and $\hat D f(x):=f(qx+\omega)$ is the dilatation operator.
One can prove that $\phi(x)$ can be at most quadratic polynomial and $\psi(x)$, at most linear one:
\be
\phi(x)=\phi_2 x^2+2\phi_1 x+\phi_0,\ \ \ \ \ \ \ \psi(x)=2\psi_1 x+\psi_0
\ee
Then the eigenvalues are,
\be
\lambda_n=[n]_q \Big(\dfrac{\phi_2}{q^n}[n-1]_q + 2\psi_1 \Big), \ \  \text{where} \ \
[n]_q := \dfrac{1 - q^n}{1 - q}.
\ee
In this case, the Rodriguez formula is still satisfied. For the monic orthogonal polynomial  is looks as
\be
{\cal P}_n(x)={K_n\over \mu'(x)}\hat H\hat D^{-n}\mu'(x)\prod_{i=1}^n\Big(\hat D^{-i-1}\phi(x)\Big)
\ee
where $d\mu(x)=\mu'(x)dx$ is the orthogonality measure, and
\be
K_n=q^n\prod_{i=1}^n{[i]\over\lambda_n-\lambda_{n-i}}
\ee
In the limit to the classical orthogonal polynomials $q\to 1$, $\omega\to 0$, one obtains the standard Rodriguez formula
\be
{\cal P}_n(x)={K_n\over \mu'(x)}{d^n\over dx^n}\Big(\mu'(x)\phi(x)^n\Big)
\ee
Note that the weight function $\mu'(x)$ satisfies the equation (which follows from the Pearson operator equation \cite{KLS})
\be
\hat H\left(\Big(D^{-k}\mu'(x)\Big)\Big(\hat D^{-k-1}\phi(x)\Big)\right)=q^{-k}\Big(\hat D^{-k+1}\mu'(x)\Big)\Big(\hat D^{-k}\psi(x)\Big)
\ee

\subsubsection*{Example. The Rodriguez formula for the Askey-Wilson polynomials}

In the case of the Askey-Wilson polynomials (\ref{AWpol}), the difference equation looks as
\be\label{AWdeq}
(q^{-n}- 1)(1-abcdq^{n-1})P^{AW}_n(z) = A(z)P^{AW}_n(qz)-\Big[A(z)+A(z^{-1})\Big] P^{AW}_n(z)+ A(z^{-1})P^{AW}_n(q^{-1}z) ,
\ee
where $z = e^{i\theta}$ and
\be
A(z) := {(1-az)(1-bz)(1-cz)(1-dz)\over (1-z^2)(1 - q z^2)}.
\ee
On the other hand, the Rodriguez formula is
\be\label{Rod}
\mu'(x; a, b,c, d;q) P^{AW}_n(x)=\Big({q-1\over 2}\Big)^n q^{n(n-1)/4}{\hat D}_q^n \mu'(x;a\sqrt{q^n},b\sqrt{q^n},c\sqrt{q^n},d\sqrt{q^n};q)
\ee
$\hat D_q = {\hat H}_q|_{\omega=0}$ and $w = \mu'$ is the orthogonality weight(\ref{AWmu}).

\section{Hypergeometric orthogonal polynomials}
\setcounter{equation}{0}
\subsection{Hypergeometric polynomials}

In this section, we are going to demonstrate that there is no immediate generalization of the orthogonal polynomials to the hypergeometric functions higher than $_4\phi_3$. That is, for an ansatz generalizing the AW polynomial, we check the 3-term relation and find that it is fulfil at most for the Askey-Wilson polynomials. Specifically, we consider two different ansatz{\bf es}: polynomials in the hypergeometric variable $z$ and in parameters of hypergeometric functions at constant $z$.

$\bullet$ Polynomials of the first type are of the following $q$-hypergeometric form
\be
\label{vcp}
^IQ_n^{(r,s)}(B,\{b_i\},\{c_i\};x) :=
{_{r}\phi_s}\left(\begin{array}{cc}
	q^{-n}, & Bq^{n},b_1,\ldots, b_{r-2} \\ &c_1,\ldots,c_s
\end{array}\Big| \,z=x\right),
\ee
where $\{b_i\}_{i=1}^{r-2}$ and $\{c_i\}_{i=1}^{s}$ are arbitrary parameters, which may depend on $n$ and $x$, while the parameter $B$ is independent of $n$ and $x$.

$\bullet$ Polynomials of the second type are $(x := \cos \theta)$,
\be
^{II}Q_n^{(r,s)}(a,B,\{b_i\},\{c_i\};x) :=
{_{r}\phi_s}\left(\begin{array}{cc}
	q^{-n}, ae^{i\theta}, ae^{-i\theta},& Bq^{n},b_1,\ldots, b_{r-4} \\ &c_1,\ldots,c_s
\end{array}\Big| \,z=q\right), \\
= \sum_{k=0}^n \frac{(q^{-n};q)_kh(x|a)_k\cdot (Bq^{n};q)_k\prod_{i=1}^{r-4} (b_i;q)_k}
{(q;q)_k\prod_{i=1}^s (c_i;q)_k}\cdot q^k\left[(-)^kq^{k(k-1)/2}\right]^{1+s-r},
\label{Polrs}
\ee
with
\be
h(x|a)_k := (ae^{i\theta};q)_k(ae^{-i\theta};q)_k = \prod_{j=0}^{k-1}(1-2axq^j + a^2q^{2j}),
\ee
where parameters $a$ and $B$ are independent on $n$ and $x$, while $\{b_i\}_{i=1}^{r-4}$ and
$\{c_i\}_{i=1}^{s}$ can be arbitrary.

Since one of the upper parameters in the $q$-hypergeometric functions (\ref{vcp}) and (\ref{Polrs}) is $q^{-n}$ with $n \in \mathbb{Z}_+$, these are, indeed, polynomials.

\subsection{Criteria for the 3-term relation}

Let us consider a polynomial $Q_n(x)$ of degree $n$ in variable $x$. We denote a coefficient of the polynomial in front of $x^m$ by $[Q_n]_m$. If polynomials $Q_n(x)$ satisfy the 3-term relation, then all $4\times 4$ determinants vanish for any $m$ such that $3\leq m\leq n+1$:

\be
\det_{4\times 4} \left(\begin{array}{cccc}
	 [x{\cdot}Q_n]_{m}   & [Q_{n+1}]_{m}   & [ Q_n]_{m}   & [Q_{n-1}]_{m} \\ \phantom{a}
	 [x{\cdot}Q_n]_{m-1} & [Q_{n+1}]_{m-1} & [ Q_n]_{m-1} & [Q_{n-1}]_{m-1} \\ \phantom{a}
	 [x{\cdot}Q_n]_{m-2} & [Q_{n+1}]_{m-2} & [ Q_n]_{m-2} & [Q_{n-1}]_{m-2} \\ \phantom{a}
	 [x{\cdot}Q_n]_{m-3} & [Q_{n+1}]_{m-3} & [ Q_n]_{m-3} & [Q_{n-1}]_{m-3}
\end{array}\right) = 0.
\ee

This vanishing  is not sufficient, but a necessary condition for the 3-term relation.
For polynomials (\ref{Polrs}), we denote these determinants  $D^{(r,s)}_{n,m}$.
They actually factorize on a codimension-one variety in the space of parameters
$b_1,\ldots,b_{r-4}$ and $c_1,\ldots, c_s$, and vanish altogether at a codimension-two
variety.

\paragraph{Examples.}
\begin{itemize}
\item {\bf For $(r,s)=(4,3)$}
\be
D^{(4,3)}_{2,3} \sim (B a^2 q - c_1c_2c_3) = 0 \ \ \ \ \ {\rm at} \ B=\frac{c_1c_2c_3}{a^2q},
\ee
and the same remains true for all other $D^{(4,3)}_{n,m}$.
These Askey-Wilson polynomials are indeed orthogonal.
\item {\bf For $(r,s)=(5,4)$}
\be
D^{(5,4)}_{2,3} \sim \prod_{i=2}^{5} (Bq^i - 1)\prod_{j=0}^1 (b_1q^j-1) \\
D^{(5,4)}_{7,8} \sim \prod_{i=6}^{15} (Bq^i - 1)\prod_{j=0}^6 (b_1q^j-1) \\
\ee
but $b_1$ can not be equal to $q^{-k}$ with $k<n$, because it reduces the degree of the polynomial $Q_n(x)$. Thus, there are no counterparts of the Askey-Wilson polynomials in this case.
\end{itemize}

\subsection{Conjecture in generic case}

In general we expect the following pattern for the $q$-hypergeometric polynomials (\ref{vcp})-(\ref{Polrs}):

\bigskip

$$
\begin{array}{|c|c|c|}
\text{type of }Q(x), & \text{name in q-Askey scheme} & (r,s) \ \text{for which the 3-term}\\
\text{parameters}&&\text{relation is satisfied}\\
&&\\
\hline
&&\\
\hbox{type I} & \text{little q-Jacobi} & (2,1) \\
&&\\
\hbox{type I}, B=\dfrac{\tilde B}{xq^{n-1}}& \text{big q-Laguerre} & (2,1) \\
&&\\
\hbox{type I}, b_1=x& \text{big q-Jacobi} & (3,2) \\
&&\\
\hbox{type I}, b_1=\dfrac{1}{x} \dfrac{c_1c_2}{Bq}& \text{equiv. to big q-Jacobi} & (3,2), \ \text{see Remark}  \\
&&\\
\hbox{type II} & \text{Askey-Wilson} & (4,3) \\
 && \\
& & \\
&\ldots & \ldots
\end{array}
$$

\paragraph{Remark.} These polynomials are equivalent to the big q-Jacobi ones with the help of a particular case of Sear's identity (see \cite[eq.(3.2.2)]{GaRa}). We write down the corresponding three-term recurrence relation, because we did not find it in the literature:
\begin{footnotesize}
\be
\label{ttr}
A_n Q_{n+1} - \left( B_n x - C_n \right) Q_n - D_n Q_{n-1} = 0, \\ \\
A_n = b{q}^{2\,n+2} \left( {q}^{4\,n-2}b-1 \right)  \left( {q}^{2\,n}b-1 \right)  \left( {q}^{2\,n}c_{{1}}-1 \right)  \left( {q}^{2\,n}c_{{2}}-1 \right), \\
B_n =  b \left( {q}^{4\,n-2}b-1 \right)  \left( {q}^{4\,n}b-1 \right)  \left( {q}^{4\,n+2}b-1 \right), \\
C_n = {q}^{2\,n} \left( {q}^{4\,n}b-1 \right) \cdot   \Big( {q}^{4\,n+2}{b}^{2}+{q}^{4\,n}b \left( bc_{{1}}{+}bc_{{2}}{+}c_{{1}}c_{{2}} \right) -{q}^{2\,n+2}b \left( b{+}c_{{1}}{+}c_{{2}} \right)    - {q}^{2\,n}b \left( c_{{1}}c_{{2}}{+}b{+}c_{{1}}{+}c_{{2}} \right) -{q}^{2\,n-2}bc_{{1}}c_{{2}}+{q}^{2}b+bc_{{1}}{+}bc_{{2}}{+}c_{{1}}c_{{2}} \Big),  \\
D_n = {q}^{n} \{q\}  [n] \left( {q}^{4\,n+2}b-1 \right)  \left( {q}^{2\,n}b-c_{{1}} \right)  \left( {q}^{2\,n}b-c_{{2}} \right).
\ee
\end{footnotesize}

\bigskip

Thus, with the help of determinant criteria, one can check what type of recurrence relations emerge for polynomials.
For the polynomials of first type, we have the following recurrence relations
$$
\begin{array}{|c|c|}
\hline
(r,s) & \text{number of terms}\\
&\text{in recurrence relation}\\
&\\
\hline
&\\
(2,1) & 3 \\
(3,2) & 5 \\
(4,3) & 5 \\
(5,4) & 7 \\
(6,5) & 7 \\
(7,6) & >7 \\
\hline
\end{array}
$$

In order to form a similar table for the polynomials of the second type, one also needs to check if the stability condition for the $q$-hypergeometric function is satisfied. The stability condition for the $q$-hypergeometric function (\ref{qhg}) reads
\be
z \prod_{i=1}^r a_i = \prod_{j=1}^s b_j,
\label{staco}
\ee
Then, we get the following table for the polynomials of the second type:
$$
\begin{array}{|c|c|c|}
\hline
(r,s) & \text{number of terms in recurrence relation} & \text{imposed stability condition}\\
&&\\
\hline
&&\\
(3,2) & 3 & no \\
(4,3) & 3 & yes \\
(5,4) & 5 & no \\
(6,5) & 5 & yes \\
(7,6) & 7 & no \\
(8,7) & 7 & yes \\
\hline
\end{array}
$$

It is clear that if one obtains an $s$-term relation for polynomials $_{s+1}\phi_s(\dots|z=q)$ under the stability condition, one can put $a_r=b_s=0$ and obtain polynomials $_{s}\phi_{s-1}(\dots|z=q)$ without the stability condition. This is precisely what we observe in the table above. Condition (\ref{staco}) is the one which allows us to substitute
all $(a)_n$ and $(b)_n$ in the definition of hypergeometric series by q-numbers $\frac{[a+n-1]!}{[a-1]!}$ and $\frac{[b+n-1]!}{[b-1]!}$ without any extra powers of $q$.

Thus, we see that,\\ \fbox{\parbox{17cm}{among the hypergeometric polynomials, the "biggest" available example of the 3-term relations and, hence, of the orthogonal polynomials is the Askey-Wilson polynomials.}}

\subsection{Orthogonal polynomials made from higher $q$-hypergeometric functions.}

Thus, there are no higher $q$-hypergeometric functions that are orthogonal polynomials. However, their linear combinations can easily form a system of orthogonal polynomials. We conclude this section with a simplest example of this kind, the Ismail-Rahman polynomials \cite{IR}. These are the systems of polynomials $\mathfrak{z}_n^{(\alpha,\beta)}(x,\vec t;q)$ of degree $n$ in $x=1/2(z+z^{-1})$
parameterized by parameters $(\alpha,t_1,t_2,t_3,t_4)$ and $\beta=0,1$:
\be
\mathfrak{z}_n^{(\alpha,\beta)}(x,\vec t;q):={zt_1^{1-2\alpha}\over 1-t_1t_2t_3t_4q^{2\alpha-2}}
{(q^\alpha ,t_1z;q)_\infty\over (t_2t_3t_4/z,t_1t_2t_3t_4q^{\alpha-1};q)_\infty}{\prod_{2\le j<k\le 4} (t_jt_k,t_jt_kq^{\alpha-1};q)_\infty\over \prod_{k=2}^4(t_1t_kq^{\alpha},t_kz;q)_\infty }\times\\
\times
\left[\Big(s_{\alpha-1}(z)-\beta s_\alpha(z)\Big)r_{n+\alpha}(z)-\Big(r_{\alpha-1}(z)-\beta r_\alpha(z)\Big)s_{n+\alpha}(z)\right],
\ee
where
\be
r_\alpha(z):={(t_2t_3t_4q^\alpha/z;q)_\infty\over (t_1zq^\alpha;q)_\infty}{\prod_{k=2}^4 (t_1t_kq^\alpha;q)_\infty\over \prod_{2\le j<k\le 4}(t_jt_kq^\alpha;q)_\infty} \Big({t_1\over z}\Big)^\alpha{_8W_7}\Big(t_2t_3t_4/(qz);t_2/z,t_3/z,t_4/z,t_1t_2t_3t_4q^{\alpha-1},q^{-\alpha};q,qz/t_1\Big),\\
s_\alpha(z):={(t_1t_2t_3t_4q^{2\alpha},t_2t_3t_4zq^{\alpha};q)_\infty\over (t_2t_3t_4zq^{2\alpha+1},q^{\alpha+1};q)_\infty}{\prod_{k=2}^4 (t_kq^{\alpha+1};q)_\infty\over \prod_{2\le j<k\le 4}(t_jt_kq^\alpha;q)_\infty} (t_1z)^{\alpha}{_8W_7}\Big(t_2t_3t_4zq^{2\alpha};t_2t_3q^\alpha,t_2t_4q^\alpha,t_3t_4q^\alpha,q^{\alpha+1},zq/t_1;q,t_1z\Big).
\ee

\section{Integral representations of orthogonal polynomials, Ward identities and conformal blocks}
\setcounter{equation}{0}

The hypergeometric series possess integral representations with multiple integrations,
which also suggest an interpretation in terms of conformal blocks.
At sub-varieties of high codimension in the space of parameters,
the hypergeometric series $_rF_s$ reduces
to those at much smaller values of $r$ and $s$
and thus can contain less integrations.
In the extreme case of the very-well-poised subvarieties (see s.\ref{3tAW} and Appendix A.2) just a single integration remains,
i.e. the corresponding conformal block contains just a single screening-charge insertion.
In this case, a natural splitting of the integration contour into three segments
provides non-trivial identities, which can be interpreted
as 3-term relations.
This gives rise to non-trivial families of orthogonal
polynomials including the Askey-Wilson/Racah as the most general known example.

In this section, we also discuss the modular transformations of the hypergeometric functions, which directly follow from their integral representations. On one hand, this is important for the issue of conformal blocks \cite{GMMF,GMMpq,Nemkov}, on the other hand, these encodes essential properties of the corresponding hypergeometric and polynomial determinant solutions to the Painl\'eve \cite{MMPVI} and discrete Painl\'eve equations \cite{JS,Sakai,MMqpain} (included into the general framework of CFT/Painl\'eve correspondence \cite{Pain,qBer}).

\subsection{Dotsenko-Fateev representations of hypergeometric functions}

Using the integral representation of the Euler Beta-function,
\be\label{beta}
B(x,y):=\int_0^1 dt\ t^{x-1}(1-t)^{y-1}={\Gamma (x)\Gamma (y)\over\Gamma (x+y)},
\ee
and the definition of the hypergeometric series (\ref{hyper}), one can directly derive the recursion relation
\be\label{mrel}
 _{r}F_{s} \left( \begin{array}{c} a_1,\ldots, a_r \\ b_1,\ldots,b_s\end{array}
\Big|\,z \right)
= \frac{\Gamma(b_s)}{\Gamma(a_{r })\Gamma(b_s-a_r)}
\int_0^1 dt\,t^{a_{r}-1}(1-t)^{b_{s}-a_{r}-1}\phantom{.}
 _{r-1}F_{s-1} \left( \begin{array}{c} a_1,\ldots, a_{r-1} \\ b_1,\ldots,b_{s-1}\end{array}
\Big|\,tz \right)
\ee
which implies that there is an integral representation with just $min(r,s)$ integrals,
which, however, does not explicitly respect the $S_r\otimes S_s$ symmetry between the
parameters $a_i$ and $b_i$ of the hypergeometric function.
In particular,
\be
 _{s+1}F_{s} \left( \begin{array}{c} a_0,a_1,\ldots, a_s \\ b_1,\ldots,b_s\end{array}
\Big|\,z \right)
=
\left(\prod_{i=1}^s \frac{\Gamma(b_i)}{\Gamma(a_i)\Gamma(b_i-a_i)} \int_0^1
 t_i^{a_i-1}(1-t_i)^{b_i-a_i-1} dt_i \right)
\cdot \overbrace{(1-t_1\ldots t_s z)^{-a_0}}^{^{\sum_k \frac{(a_0)_k}{k!}\cdot(t_1\ldots t_s z)^k}}
\label{polintrep}
\ee
or, after the change of variables $t_i\to t_i/t_{i+1}$,
\be
 _{s+1}F_{s} \left( \begin{array}{c} a_0,a_1,\ldots, a_s \\ b_1,\ldots,b_s\end{array}
\Big|\,z \right) =
\left(\prod_{j=1}^s \frac{\Gamma(b_j)}{\Gamma(a_j)\Gamma(b_j-a_j)}
\int_0^{t_{j+1}} dt_j\,t_j^{a_{j}-a_{j-1}-1}
\left(1-\frac{t_j}{t_{j+1}}\right)^{b_{j}-a_{j}-1}\right)\cdot
t_1^{a_0}(1-t_1 z)^{-a_0}.
\ee
where we put $t_{s+1}=1$.

These representations can be associated with the free field representation of the conformal blocks \cite{DF,confmamo,DF4} (see \cite{DF5} for the $q$-deformation), see s.\ref{DFcb}.

\subsubsection{Example: Ward and bilinear identities}

As soon as hypergeometric function have integral representations, various relations between these functions can be interpreted as Ward identities, as explained in \cite{MMqpain}. The simplest examples of such relations is
\be
_2F_1\left(\begin{array}{c}a,b\\c\end{array}\Big|\,x\right) = \frac{\Gamma(c)}{\Gamma(b)\Gamma(c-b)}
\int_0^1 (1-xt)^{-a}t^{b-1}(1-t)^{c-b-1}dt
= (1-x)^{c-a-b} \cdot {_2F_1}\left(\begin{array}{c}c-a,c-b\\c\end{array}\Big|\,x\right),
\label{Fid1}
\ee
which can be obtained by the change of integration variable $t\longrightarrow \frac{1-t}{1-xt}$.

Another example is provided by the replace $t\longrightarrow 1-t$ and leads to the Euler transformation formula
\be\label{Euler}
_2F_1\left(\begin{array}{c}a,b\\c\end{array}\Big|\,x\right) = \frac{\Gamma(c)}{\Gamma(b)\Gamma(c-b)}
\int_0^1 (1-xt)^{-a}t^{b-1}(1-t)^{c-b-1}dt
=(1-x)^{-a}\phantom{.}_2F_1\left(\begin{array}{c}a,c-b\\c\end{array}\Big|\,{x\over x-1}\right).
\ee

Since
\be
(1-x)^{-\alpha} = \sum_{n=0}^\infty \frac{(\alpha)_n}{n!}\,x^n
\ee is also a hypergeometric series, one can also interpret (\ref{Fid1}) as a kind
of bilinear identity \cite{BI}.
In the lowest orders, it is trivial:
\be
1 + \frac{ab}{c}\, x + \ldots = \Big(1+(a+b-c)x+\ldots\Big)
\Big(1+\frac{(c-a)(c-b)}{c}\,x + \ldots\Big),
\ee
i.e. $ab = (c-a)(c-b) + c(a+b-c)$ etc,
but in general it looks more interesting:
\be
\frac{(a)_n(b)_n}{n!\,(c)_n} =
\sum_{j=0}^n \frac{(c-a)_j(c-b)_j}{(c)_j}\cdot\frac{(a+b-c)_{n-j}}{j!(n-j)!}
= \frac{(c-a-b)_n}{n!}\sum_{j=0}^n \frac{(-n)_j(c-a)_j(c-b)_j}{(c)_j(c-a-b-n+1)_j\,j!},
\ee
i.e.
\be
_3F_2\left(\begin{array}{c}-n,c-a,c-b\\c,c-a-b-n+1\end{array}\Big|\,x=1\right)
 = \frac{(a)_n(b)_n}{(c)_n(a+b-c)_n},
\label{3F2fact}
\ee
which is nothing but Pfaff-Saalsch\"utz's identity, \cite[s.1.7]{GaRa}.

\subsubsection{Hypergeometric vs. modular transformations\label{DFcb}}

Now let us discuss the relation with conformal blocks. Consider the conformal block at $c=1$, $\alpha_4=-\alpha_1-\alpha_2-\alpha_3-1$  \cite{MMPVI}:
\be
B(\alpha_i,\alpha;x)\Big|_{\alpha=\alpha_1+\alpha_2+1}= x^{2\alpha_1\alpha_2+2\alpha_1+2\alpha_2+1}(1-x)^{2\alpha_2\alpha_3}\ {_2F_1}\left(\begin{array}{c}-2\alpha_3,2\alpha_1+1\\ 2\alpha_1+2\alpha_2+2\end{array}\Big|\,x\right),\\
B(\alpha_i,\alpha;x)\Big|_{\alpha=\alpha_1+\alpha_2}=x^{2\alpha_1\alpha_2}(1-x)^{2\alpha_2\alpha_3} \ {_2F_1}\left(\begin{array}{c}-2\alpha_1-2\alpha_2-2\alpha_3-1,-2\alpha_2\\ -2\alpha_1-2\alpha_2\end{array}\Big|\,x\right),
\ee
where $\alpha_i$, $i=1,2,3,4$ parameterize the external conformal dimensions, $\Delta_i=\alpha_i^2$, while $\alpha$, the intermediate one, $\Delta=\alpha^2$.

Then, (\ref{Fid1}) can be rewritten as
\be\label{MT1}
B(\alpha_1,\alpha_2,\alpha_3,\alpha_4,\alpha;x)=(1-x)^{2(\alpha_1+\alpha_2)(\alpha_1+\alpha_3)-2\alpha_4}B(\alpha_2,\alpha_1,\alpha_4,\alpha_3,\alpha;x).
\ee
It is nothing but a result of modular
transformation of the conformal block \cite{AGTsu3}. Indeed, the generic four-point
function is \cite{BPZ,CFT}
\be\label{4V}
\langle V_{\Delta_1}(z_1)V_{\Delta_2}(z_2)V_{\Delta_3}(z_3)V_{\Delta_4}(z_4)
\rangle =\prod_{i>j}^4 (z_i-z_j)^{d_{ij}}(\bar z_i-\bar z_j)^{\bar d_{ij}}G(x,\bar x),
\ee
where $d_{ij}$ are determined by the equations
\be
\sum_{j\ne i}d_{ij}=2\Delta_i
\ee
$x={(z_1-z_2)(z_3-z_4)\over
(z_1-z_3)(z_2-z_4)}$ and similarly for the complex conjugated part.
$G(x,\bar x)$ is a bilinear combination of the conformal blocks
$B(\{\Delta_i\},\Delta,,c;x)$ and
$B(\{\bar\Delta_i\},\bar\Delta, c;\bar x)$.
Due to the projective invariance, one can choose three of these four
points arbitrarily. If choosing $z_4=\infty$, $z_3=1$ and $z_1=0$,
one obtains $z_2=x$, and the 4-point correlator (\ref{4V}) becomes
\be
(1-x)^{d_{32}}\bar x^{\bar d_{21}}(1-\bar x)^{\bar d_{32}}x^{d_{21}}G(x,\bar x).
\ee
The simultaneous interchange of
external lines $\vec\alpha_1 \leftrightarrow \vec\alpha_2$
and $\vec\alpha_3 \leftrightarrow \vec\alpha_4$ in formula (\ref{4V})
does not change $x$ and $\bar x$
and leads to the factor of $(1-x)^{d_{14}}$ instead of $(1-x)^{d_{23}}$,
which is exactly the factor $(1-x)^{2(\alpha_1+\alpha_2)(\alpha_1+\alpha_3)-2\alpha_4}$ in (\ref{MT1}):
\be
d_{23}-d_{14}=\Delta_2+\Delta_2-\Delta_1-\Delta_4=2(\alpha_1+\alpha_2)(\alpha_1+\alpha_3)-2\alpha_4
\ee
Similarly, the interchange of only external lines $\vec\alpha_1 \leftrightarrow \vec\alpha_2$ in formula (\ref{4V}) changes $x\to x/(x-1)$ and analogously $\bar x$ and is described by the Euler transformation formula (\ref{Euler}).

Note that another modular transformation, which replaces $x\to 1-x$, is far less trivial \cite{AG,MS,GMMF} and
corresponds in terms of hypergeometric functions to the identity
\be
{_2F_1}\left(\begin{array}{c}a,b\\c\end{array}\Big|\,x\right)={\Gamma(c)\Gamma(c-a-b)\over\Gamma(c-a)\Gamma(c-b)}\ {_2F_1}\left(\begin{array}{c}a,b\\a+b-c+1\end{array}\Big|\,1-x\right)+\\
+(1-x)^{c-a-b}{\Gamma(c)\Gamma(a+b-c)\over\Gamma(a)\Gamma(b)}\
 {_2F_1}\left(\begin{array}{c}c-a,c-b\\c-a-b+1\end{array}\Big|\,1-x\right)
\ee
In this case, this transformation maps the conformal block to another one with different conformal dimensions. One can apply the same transformation again and obtain a sequence of conformal blocks with different conformal dimensions. This sequence is infinite unless the conformal dimensions are specifically constrained. If they are properly constrained, the sequence becomes finite, the modular transformation becomes a linear transformation in the finite-dimensional space of the corresponding conformal blocks, and this is exactly the case of minimal models \cite{BPZ,DF}.

\subsubsection{Integral representation of the very classical polynomials}

\paragraph{Jacobi polynomials.}
One could try to use a particular example of formula (\ref{polintrep}) at $s=1$ (i.e. the Gaussian hypergeometric function)
\be
 _2F_1 \left( \begin{array}{c} a,b\\ c\end{array}
\Big|\,z \right)
=
\frac{\Gamma(c)}{\Gamma(b)\Gamma(c-b)} \int_0^1
dt\ t^{b-1}(1-t)^{c-b-1}(1-tz)^{-a},
\ee
and the definition of the Jacobi polynomials (\ref{jac}) in order to obtain an integral representation of these latter
\be
P_n^{(a,b)}(z)={(a+n)!\over 2^nn!(a+b+n)!\Gamma (-b-n)}\int_0^1 dt\ t^{a+b+n}(1-t)^{-b-n-1}(zt-t+2)^n.
\ee
Unfortunately, this integral is never defined at all natural $n$ at once. However, we can proceed in a more tricky way \cite{Askey74}. First of all, one can obtain Bateman's integral
\be\label{Bate}
 _2F_1 \left( \begin{array}{c} a,b\\ c\end{array}
\Big|\,z \right)
={1\over B(c',c-c')}\int_0^1 dt\ t^{c'-1}(1-t)^{c-c'-1}\phantom{.} _2F_1 \left( \begin{array}{c} a,b\\ c'\end{array}
\Big|\,zt \right),
\ee
just from the definition of the hypergeometric function as a series (\ref{hyper}) and evaluating the integrals with the help of (\ref{beta}). Now, combining this formula with (\ref{jac}) and with the Euler transformation (\ref{Euler}), one immediately arrives at
\be
P^{(a,b)}_n(z)={2^{a-b}\over B(a-b,b+n)}{(1+z)^{b+n+1}\over (1-z)^{a}}\int_z^1 dt\ (1-t)^b(1+t)^{-a-n-1}(t-z)^{a-b-1}P_n^{(b,b)}(t).
\ee
Making the substitution $\xi^2=\displaystyle{(1-t)(1+z)\over (1+t)(1-z)}$, one obtains the integral formula that expresses the Jacobi polynomials $P^{(a,b)}_n(z)$ through $P^{(b,b)}_n(z)$, which are equal to {\bf the Gegenbauer (or ultraspherical) polynomials} up to a normalization:
\be
P_n^{(a,b)}(z)={(-1)^{n}2^{1-n}\over B(a-b,b+n+1)}(1+z)^2\int_0^1 d\xi\ \xi^{2b+1}\Big((z-1)\xi^2-z-1\Big)^n(1-\xi^2)^{a-b-1}P_n^{(b,b)}\left({(1-z^2)(\xi^2-1)\over \xi^2(z-1)-z-1}\right).
\ee
$P^{(b,b)}_n(z)$ have the integral representation\footnote{This can be checked by expanding (\ref{Geir}) into the Newton binomial and using the integrals
\be
\int_0^\pi d\theta\sin^{2b}\theta\cos^k\theta=\left\{
\begin{array}{ll}
0&\hbox{for }k=2n+1,\\
4^b B\Big(b+{1\over 2},b+{1\over 2}\Big)(b+1)_n{\Gamma \Big(n+{1\over 2}\Big)\over\sqrt{\pi}}&\hbox{for }k=2n.
\end{array}\right.\nonumber
\ee
This leaves us with a sum of terms $z^{2n-2k}(1-z^2)^k$ which, upon with the substitution $(1-z)/2=y$, i.e. $z=1-2y$, $1-z^2=4y(1-y)$ can be expanded into degrees of $y$. Now one can check that the coefficient in front of each $y_p$ coincides with the corresponding coefficient of $P^{(b,b)}_n(z)$ as a hypergeometric series (\ref{hyper}) of $y=(1-z)/2$.
}
\be\label{Geir}
P_n^{(b,b)}(z)={\Gamma (b+n+1)\over n!\Gamma \Big(b+{1\over 2}\Big)\Gamma\Big({1\over 2}\Big)}\int_0^\pi d\theta\ \sin^{2b}\theta\Big[z+i\sqrt{1-z^2}\cos\theta\Big]^n.
\ee
Hence, one finally obtains an integral representation of the Jacobi polynomials \cite{Koorn72}
\be
P_n^{(a,b)}(z)={2^{1-n}\Gamma (a+n+1)\over n!\Gamma (a-b)\Gamma \Big(b+{1\over 2}\Big)\Gamma\Big({1\over 2}\Big)}\times\\
\times\int_0^1 d\xi\int_0^\pi d\theta\
(1-\xi^2)^{a-b-1}\xi^{2b+1}\sin^{2b}\theta \Big[1+z-(1-z)\xi^2+2i\sqrt{1-z^2}\ \xi\cos\theta\Big]^n.
\ee
This integral exists at $a>b>-1/2$. At a wider range of parameters $a,b>-1/2$, there is another, more symmetric double integral representation for the Jacobi polynomials, \cite{BMDK}\footnote{The simplest way to prove this formula is to expand $(\ldots )^{2n}$ in the integrand into the Newton binomial, use (\ref{beta}) and then use the Newton binomial expansion for $(1-y^2)^{k}$ again. Then, using
\be
\sum_{l\le k\le n}{k!\Gamma\Big(k+{1\over 2}\Big)\Gamma\Big(n-k+{1\over 2}\Big)\over (k-l)!(2k)!(2n-2k)!\Gamma (k+a+1)\Gamma (n-k+b+1)}=
{4^{-n}\pi\ \Gamma (a+b+2n-l+1)\over\Gamma (a+n+1)\Gamma (b+n+1-l)\Gamma(a+b+n+1)(n-l)!}.\nonumber
\ee
one easily checks that the coefficients of expansion coincide with the coefficients of series (\ref{jac}) in $(1-x)/2=y^2$.}
\be
P_n^{(a,b)}(1-2y^2)={(-4)^n\over\pi (2n)!}{\Gamma (a+n+1)\Gamma (b+n+1)\over\Gamma\Big(a+{1\over 2}\Big)\Gamma\Big(b+{1\over 2}\Big)}
\times\\
\times\int_{-1}^1 d\xi_1\int_{-1}^1 d\xi_2\ \Big[y\xi_1\pm i\sqrt{1-y^2}\ \xi_2\Big]^{2n}(1-\xi_1^2)^{a-1/2}(1-\xi_2^2)^{b-1/2}.
\ee

In the following, we shall derive integral representations of the Laguerre and Hermite polynomials, which is of relevance when we discuss the use of orthogonal polynomials in the study of the matrix
model in sec.8.

\paragraph{Laguerre polynomials.}

Another particular application of formula (\ref{mrel}) is to the case of the degenerate hypergeometric function $_1F_1$:
\be\label{mrel-1}
 _{1}F_{1} \left( \begin{array}{c} a\\ b\end{array}
\Big|\,z \right)
= \frac{\Gamma(b)}{\Gamma(a)\Gamma(b-a)}
\int_0^1 dt\,t^{a-1}(1-t)^{b-a-1}\phantom{.}
 _{0}F_{0} (tz)=\frac{\Gamma(b)}{\Gamma(a)\Gamma(b-a)}
\int_0^1 dt\, e^{tz}t^{a-1}(1-t)^{b-a-1}.
\ee
One may try to use the definition of the Laguerre polynomials as degenerate hypergeometric functions (\ref{Lag}) in order to obtain their integral representation
\be
L^{(a)}_n(z) \ = \ \frac{1}{\Gamma(-n)\Gamma(1-n)}\int_0^1 dt\, e^{tz}t^{-n-1}(1-t)^{n+a}.
\ee
However, this formula requires a regularization because of the zeroes at natural $n$ due to $\Gamma (-n)$ in the denominator and divergency of the integral at zero. Hence, we use another formula,
\be\label{Bes}
\phantom{.}_{1}F_{1} \left( \begin{array}{c} a \\ b\end{array}
\Big|\,z \right)={\Gamma (b)\over\Gamma (b-a)}e^zz^{(1-b)/2}\int_0^\infty dt\ e^{-t}t^{(b-1)/2-a}J_{b-1}(2\sqrt{tz}),
\ee
where $J_b(z)$ is the Bessel function of order $b$. This formula is proved by expanding the Bessel function into the series
\be
J_b(z)=\Big({z\over 2}\Big)^b\ \sum_{k=0}^\infty {(-1)^k\over k!\Gamma (k+b+1)}\Big({z\over 2}\Big)^{2k},
\ee
and calculating the integrals with the help of the first formula in (\ref{Gamma}). The series obtained coincides with the degenerate hypergeometric series at the l.h.s. Using now the definition of the Laguerre polynomials (\ref{Lag}), one finally obtains from (\ref{Bes})
\be\label{irLag}
L^{(a)}_n(z)={1\over n!}e^zz^{-a/2}\int_0^\infty dt\ e^{-t}t^{n+a/2}J_a(2\sqrt{tz}).
\ee

\paragraph{Hermite polynomials.}

In this case, we can use the definition (\ref{herm}) and note that
\be
H_n(z)=\left\{
\begin{array}{cl}
(-4)^mm!\ L_m^{(-{1\over 2})}(z^2)&n=2m,\\
2(-4)^mm!\ zL_m^{({1\over 2})}(z^2)&n=2m+1.
\end{array}
\right.
\ee
Then, the integral representation is read off from (\ref{irLag}):
\be\label{irHer}
H_n(z)={2^{n+1}\over\sqrt{\pi}}e^{z^2}\int_0^\infty dt\ e^{-t^2}t^n\cos\Big( 2zt-{n\pi\over 2}\Big)={(2i)^n\over\sqrt{\pi}}\int_{-\infty}^\infty dt\
t^ne^{(z+it)^2},
\ee
where we have taken into account that
\be
J_{\pm {1\over 2}}(z)=\sqrt{2\over \pi z}\sin \Big(z+{\pi (1\pm 1)\over 4}\Big),
\ee
and changed the integration variable $t\to t^2$.

\subsection{A single-integral representation for the AW type polynomials}

This is another integral representation for the AW type polynomials, which requires just one integration.
Defining the Jackson integral
\be\label{Jack}
\int_a^b f(t)\,d_qt = (1-q)  \sum_{n=0}^\infty q^n \Big[ b f(bq^n)- a f(aq^n) \Big],
\ee
the $q$-hypergeometric function at $z=q$ is
\be
_r\phi_s\left(\begin{array}{c} a_1,\ldots, a_r\\ b_1,\ldots, b_s\end{array}\Big| z\right)
= \sum_{n=0}^\infty \frac{(a_1;q)_n\ldots (a_r;q)_n}{(b_1;q)_n\ldots (b_s;q)_n}
\,\frac{z^n}{(q;q)_n} \ \longrightarrow   \\
\stackrel{z=q}{\longrightarrow}\
\frac{1}{1-q}\cdot \frac{(a_1,\ldots,a_r;q)_\infty}{(b_1,\ldots,b_s,q\,;q)_\infty}
\int_0^1 d_qt \ \frac{(b_1t,\ldots,b_st,qt;q)_\infty}{(a_1t,\ldots,a_rt;q)_\infty},
\label{zqintrep}
\ee
i.e. the $q$-hypergeometric series at fixed $z=q$ is always a single (Jackson) integral. This gives us an integral representation for any orthogonal polynomials of the second type.

\subsubsection{Example: 3-term relation and integral representation for the Askey-Wilson polynomials 
\label{3tAW}}

Consider the very-well-poised hypergeometric series $_8W_7$ (\ref{W}) at a special point, where it has an integral representation
with just one integration:
\be\label{intf}
_8W_7(a;b,c,d,e,f\Big|q,{a^2q^2\over bcdef})={aq-def\over adefq(1-q)}
{(aq,d,e,f,aq/bc,aq/de,aq/df,aq/ef;q)_{\infty}\over
(q,aq/b,aq/c,aq/d,aq/e,aq/f,def/aq,aq/def;q)_{\infty}}\times\\
\times\int_{aq}^{def}{(t/a,qt/def,aqt/bdef,aqt/cdef;q)_{\infty}\over
(t/de,t/df,t/ef,aqt/bcdef;q)_{\infty}}d_qt.
\ee
Dividing the integration segment into two pieces:
\be
\int_{aq}^{def}=\int_{aq}^{bdef/a}+\int_{bdef/a}^{def},
\ee
one obtains the Bailey three-term identity:
\be\label{Bailey}
{1\over a}{(aq/d,aq/e,aq/f,q/ad,q/ae,q/af;q)_{\infty}\over (qa^2,ab,ac,b/a,c/a;q)_{\infty}} {_8W_7}(a^2;ab,ac,ad,ae,af|q,q^2/abcdef)+
(a\leftrightarrow b)+(a\leftrightarrow c)=0.
\ee
It remains to note that
\be
_8W_7(a,b,c,d,e,q^{-n}|q,a^2q^{2+n}/bcde)={(aq,aq/de;q)_n\over (aq/d,aq/e)_n} {_4\phi_3}\left(\begin{array}{c}q^{-n},d,e,aq/bc\\
aq/b,aq/c,deq^{-n}/a \end{array}\Big|\,q,q\right).
\ee
Hence, the Askey-Wilson polynomials,
\be\label{AW}
P_n(x=\cos\theta;a,b,c,d;q)=(ab,ac,ad;q)_n a^{-n}{_4\phi_3}\left(\begin{array}{c}q^{-n},abcdq^{n-1},ae^{i\theta},ae^{-i\theta}\\
ab,ac,ad \end{array}\Big|\,q,q\right),
\ee
satisfy the three-term recurrence relation that follows from (\ref{Bailey}) and, hence, they are orthogonal polynomials.

There are three technical points behind this line: (i) integral representation (\ref{intf}), the proof is in \cite[s.2.10]{GaRa}; (ii) the relation between $_8W_7$ and $_4\phi_3$, the proof is in \cite[s.2.5]{GaRa} and (iii) the orthogonality measure, the manifest expression being available from \cite[s.7.5]{GaRa}.

From these formulas, we also immediately obtain the integral representation for the Askey-Wilson polynomials:
\be
P_n^{AW}(x=\cos\theta|a,b,c,d)={2id\over q(1-q)}{(ab,ac,bc;q)_n\over (ab,ac,bc,q;q)_\infty h(x|d)w(x|a,b,c,d)}\times\\
\times
\int_{qe^{i\theta}\over d}^{qe^{-i\theta}\over d} d_qu\ {(due^{i\theta},due^{-i\theta},abcdu/q;q)_\infty\over (adu/q,bdu/q,cdu/q;q)_\infty}{(q/u;q)_n\over
(abcdu/q;q)_n}\Big({du\over q}\Big)^n.
\ee

\subsubsection{Mellin-Barnes representation}

Eq.(\ref{zqintrep}) with arbitrary $z$ is a counterpart of the Mellin-Barnes representation
for the classical ($q=1$) hypergeometric series:
\be
{\prod_{k=1}^{r}\Gamma(a_k)\over\prod_{k=1}^{s}\Gamma(b_k)}\phantom{1}_{r}F_s(a_1,...,a_{r}; \ b_1,\ldots,b_{s}\,|\,z)={1\over 2\pi i}
\int_{-i\infty}^{i\infty}{\prod_{k=1}^{r}\Gamma(a_k+t)\over\prod_{k=1}^{s}\Gamma(b_k+t)}\Gamma(-t)(-z)^tdt.
\ee
It is extended to the $q$-hypergeometric series
in the form of an ordinary contour integral:
\be
_{s+1}\phi_s\left(\begin{array}{c} a_1,\ldots, a_{s+1}\\ b_1,\ldots, b_s\end{array}\Big| z\right)={(a_1,a_2,\ldots,a_{s+1};q)_{\infty}\over
(b_1,\ldots,b_s,q\,;q)_{\infty}}\Big(-{1\over 2\pi i}\Big)\int_{-i\infty}^{i\infty}{(q^{t+1},b_1q^t,\ldots,b_sq^t;q)_{\infty}\over
(a_1q^t,a_2q^t,\ldots,a_{s+1}q^t;q)_{\infty}}{\pi(-z)^tdt\over\sin\pi t}.
\ee

\section{Racah as hypergeometric discrete orthogonal polynomials}
\setcounter{equation}{0}
\subsection{Racah polynomials as orthogonal polynomials of discrete variable\label{Racah}}

We describe here the $q$-Racah polynomials \cite{GaRa} that are maximal discrete hypergeometric polynomials, similarly to the Askey-Wilson polynomials being maximal continuous ones. Consider the Askey-Wilson polynomials (\ref{AWpol}) without the normalizing factor and impose the requirement $ac=q^{-N}$ with some natural $N$. Then, one can define the $q$-Racah polynomials (with redefined parameters as compared with (\ref{AWpol})):
\be\label{Racahs}
\mathfrak{R}_n(j;a,b,c,N;q)={_4\phi_3}\left(\begin{array}{c}q^{-n},abq^{n+1},q^{-j},cq^{j-N}\\
aq,q^{-N},bcq \end{array}\Big|\,q,q\right)
\ee
These are polynomials of the discrete variable $x_j=q^{-j}+cq^{j-N}$ given on a finite set $j=0,1,\ldots,N$. In fact, they are also polynomials of degree $j$ in $y_n=q^{-n}+abq^{n+1}$, which is a counterpart of the duality relation (\ref{duality}) for the Askey-Wilson polynomials.

The orthogonal relations (\ref{dopr}),
\be\label{Rorth}
\sum_{j=1}^N \mathfrak{R}_n(x_j) \ \mathfrak{R}_m(x_j)\ \mu_j = ||\mathfrak{R}_n||^2 \ \delta_{m,n}
\ee
is now satisfied with the weight, $\mu_j=\mu(x_j)$, defined as
\be\label{Rmea}
\mu_j={1-cq^{2j-N}\over 1-cq^{-N}}{(cq^{-N},aq,bcq,q^{-N};q)_j\over (q,cq,cq^{-N}/a,q^{-N}/b;q)_j}(abq)^{-j}
\ee
and the norms of the polynomials are
\be\label{Rnorm}
||\mathfrak{R}_n||^2={1-abq\over 1-abq^{2n+1}}{(abq^2,1/c;q)_N\over (bq,aq/c;q)_N}{(q,bq,aq/c,abq^{N+2};q)_n\over (aq,abq,bcq,q^{-N};q)_n}(cq^{-N})^n
\ee
The parameters $a$, $b$, $c$ are such that (\ref{Rmea}) and (\ref{Rnorm}) are positive.

Note that (\ref{dual}) implies that considered as polynomials in $y_n=q^{-n}+abq^{n+1}$, the $q$-Racah polynomials are also orthogonal with the weight $1/||\mathfrak{R}_n||^2$ and the norm square $1/\mu_j$:
\be
\sum_n {\mathfrak{R}_n(x_i) \mathfrak{R}_n(x_j) \over ||\mathfrak{R}_n||^2} =
\sum_n {\mathfrak{\tilde R}_i(y_n) \mathfrak{\tilde R}_j(y_n) \over ||\mathfrak{\tilde R}_n||^2}={\delta_{i,j}\over \mu_j}.
\ee

Consider now the matrix $U$ with the matrix elements
\be
\label{ortmat}
U_{ij}\equiv {\sqrt{\mu_j}\over ||\mathfrak{R}_i||}\mathfrak{R}_i(x_j)
\ee
Then, from the orthogonality relation
\be
\sum_{j=0}^N \mathfrak{R}_i(x_j)\mathfrak{R}_k(x_j)\mu_j=||\mathfrak{R}_i||^2\delta_{i,k}
\ee
it follows that $U$ is orthogonal. The dual orthogonality relation (\ref{dual}) is the same orthogonality relation for the transposed matrix $U^t$.

\subsection{Relation between $\mathfrak{R}_n$ and $P^{AW}_n$}

From
\be
\mathfrak{R}_n(\tilde x_j) = {_4\phi_3}\left(\begin{array}{c}
q^{-n},\tilde a\tilde bq^{n+1},q^{-j},\tilde cq^{j-N}\\
\tilde aq,q^{-N},\tilde b\tilde cq \end{array}\Big|\,z=q \right)= \\
= {_4\phi_3}\left(\begin{array}{c}q^{-n},abcdq^{n-1},ae^{i\theta},ae^{-i\theta}\\
ab,ac,ad \end{array}\Big|\,z=q\right)
={a^n\over (ab,ac,ad;q)_n}\ P^{AW}_n(x|a,b,c,d)
\ee
and $\tilde x_j = q^{-j}+ \tilde cq^{j-N}$, $x = \cos\theta$
we have:
\be\label{6.9}
a^2 = \tilde c q^{-N}   \\
b = \frac{\tilde aq}{a}, \ \ c= \frac{q^{-N}}{a}, \ \ d = \frac{\tilde b\tilde c q}{a} \\
\Longrightarrow \ abcd = \frac{\tilde a\tilde b\tilde c q^{2-N}}{a^2} =
\tilde a\tilde b q^2
\ee
and
\be\label{tildex}
\tilde x_j = a \Big(aq^j+a^{-1} q^{-j}) = 2a \cos\theta = 2a x
\ee

Change of variables does not affect (just redefines) the 3-term relation. As for the measures in these two cases, let us note that the orthogonality relation (\ref{AWmeasure}) for the Askey-Wilson polynomials is singular at $ac=q^{-N}$ for the polynomials with $n\le N$ being proportional to $\displaystyle{(q^{-N};q)_n\over (q^{-N};q)_\infty}$. However, this is exactly the case of the Racah polynomials. In fact, as it was already noted in s.\ref{AW}, in the case of $|a|>1$, the orthogonality relation (\ref{AWmeasure2}) contains an additional sum as compared with (\ref{AWmeasure}) coming from pole contributions outside the unit circle
\be
\int_{-1}^1 P^{AW}_m(x)P^{AW}_n(x) w(x)dx+2\pi\sum_k  P^{AW}_m(x_k)P^{AW}_n(x_k)w_k= \frac{\delta_{m,n}}{h_n}\int_{-1}^1 w(x)dx
\ee
and the sum runs over the points $x_k=(aq^k+q^{-k}/a)/2$. Multiplying the both sides of this identity with the factor $\displaystyle{(q^{-N};q)_\infty\over (q^{-N};q)_n}$, one makes the orthogonality relation non-singular, but at $n\le N$ the first term at the l.h.s. vanishes, and one remains with the orthogonality relation for the Racah polynomials with measure (\ref{Rmea}) (one has to take into account changing the normalization of the Racah polynomials as compared with the Askey-Wilson ones), the points $x_k=(aq^k+q^{-k}/a)/2$ becoming exactly the proper Racah polynomial variables upon identification (\ref{tildex}).

\section{Generalizations of orthogonal polynomials}
\setcounter{equation}{0}

In this section, we consider various generalizations of the hypergeometric polynomials: to the multivariable polynomials and to functions instead of polynomials. In fact, there are two distinct multivariable generalizations, we consider also discrete and continuous orthogonal multivariable polynomials.

\subsection{Multivariable extensions of Askey-Wilson and Racah polynomials}

\subsubsection{$BC$ Koornwinder-Macdonald polynomials}
The Askey-Wilson (q-Racah) polynomials can be described as simplest (one-variable, or one-row) symmetric polynomials for the systems of roots of the $BC_n$ type \cite{Mac}. This has been first realized by T.Koornwinder \cite{Koor} (hence, the name Koornwinder-Macdonald polynomials). These symmetric polynomials are constructed from the monomials
\be
m_\lambda=\sum_{\mu\in G(\lambda)} z_1^{\mu_1}z_2^{\mu_2}\ldots z_n^{\mu_s}
\ee
where the sum goes over the orbit $G(\lambda)$ of the partition $\lambda=\{\lambda_1\ge\lambda_2\ge\ldots\ge 0\}$ under the action of the group $G=S_s\times \mathbb{Z}_2$ which permutes $\lambda_i$ and changes their signs. This is nothing but the $BC_s$-type Weyl group. Now defining the second order difference operator \cite{Koor}
\be\label{difop}
\hat{\mathfrak{D}}=\sum_j\left(P(z;z_j)(\hat T_j-1)+P(z;z_j^{-1})(T_j^{-1}-1)\right)
\ee
where
\be
P(z;z_j)={\prod_{a=0}^3 (1-t_az_j)\over (1-z_j^2)(1-qz_j^2)}\prod_{k\ne j}{(1-tz_jz_k)(1-tz_jz_k^{-1})\over (1-z_jz_k)(1-z_jz_k^{-1})},\ \ \ \ \ \ \
\hat T_j f(z_1,z_2,\ldots)=f(z_1,z_2,\ldots,qz_j,\ldots)
\ee
one can construct the set of $BC$-Macdonald (or Koornwinder-Macdonald) polynomials
\be
P^{KM}_\lambda(z|t_0,t_1,t_2,t_3,t)=\left(\prod_{\mu\le\lambda}{\hat{\mathfrak{D}}-E_\mu\over E_\lambda-E_\mu}\right)m_\lambda
\ee
where
\be
E_\lambda=\sum_{j=1}^s q^{-1}t_0t_1t_2t_3t^{2n-j-1}(q^{\lambda_j}-1)+t^{j-1}(q^{-\lambda_j}-1)
\ee
and $\mu\le\lambda$ is understood as $\sum_{j=1}^k(\lambda_j-\mu_j)\ge 0$ for all $k=1,\ldots,s$.
These polynomials are the eigenfunctions of the difference operator $\hat{\mathfrak{D}}$ (\ref{difop}) with the eigenvalues $E_\lambda$:
\be
\hat{\mathfrak{D}}P^{KM}_\lambda=E_\lambda P^{KM}_\lambda
\ee
Considering the case of one variable $z$, i.e. $s=1$, which is equivalent to one-line Young diagram $\lambda$ corresponding to symmetric representations, one arrives at the Askey-Wilson polynomials (\ref{AWpol}) with $z=e^{i\theta}$ and $a=t_0$, $b=t_1$, $c=t_2$, $d=t_3$:
\be
P^{KM}_{[n]}(e^{i\theta}|t_0,t_1,t_2,t_3,t)=P^{AW}_n(\cos\theta|a,b,c,d)
\ee
Note that, in this case, the $t$-dependence automatically disappears from the polynomial.

\subsubsection{Multivariable Askey-Wilson and Racah polynomials}

\paragraph{Multivariable Askey-Wilson polynomials.}
There is another multivariable generalization of the Askey-Wilson polynomials \cite{Trat,GR2,Il}:
\be
P^{AW}_{\bf n}({\bf x}|\alpha_i)=\prod_{i=1}^sP^{AW}_{n_i}\Big(x_i=\cos\theta_j\Big|\alpha_iq^{N_{i-1}},{\alpha_i\over\alpha_0^2},{\alpha_{j+1}\over\alpha_j}e^{i\theta_{j+1}},
{\alpha_{j+1}\over\alpha_j}e^{-i\theta_{j+1}}\Big)
\ee
where ${\bf n}=(n_1,\ldots,n_s)\in \Big(\mathbb{Z}_{\ge 0}\Big)^s$, ${\bf x}=(x_1,\ldots,x_s)$, $N_k:=\sum_{i=0}^k n_k$ with the conventions $N_0:=0$, $e^{i\theta_0}=\alpha_0$ and $e^{i\theta_{s+1}}=\alpha_{s+2}$.

A counterpart of the conditions $|a|<1$,$|b|<1$, $|c|<1$, $|d|<1$ for the Askey-Wilson polynomial (\ref{AWpol}) are the conditions
\be
0<|\alpha_{s+1}|<|\alpha_s|<\ldots<|\alpha_1|<\hbox{min}(1,|\alpha_0|^2),\ \ \ \ \ \ \ \ \ \ {|\alpha_{s+1}|\over |\alpha_s|}<
|\alpha_{s+2}|< {|\alpha_{s}|\over |\alpha_{s+1}|}
\ee
The orthogonality conditions in this case are
\be
\int_0^1{dx_1\over 2\pi\sqrt{1-x_1^2}}\ldots \int_0^1{dx_s\over 2\pi\sqrt{1-x_s^2}}W({\bf x}) P^{AW}_{\bf n}({\bf x})P^{AW}_{\bf m}({\bf x})=
H_{\bf n}\delta_{\bf m,n}
\ee
with the weight and norm squared given as:
\be
W({\bf x})={\prod_{i=1}^s (e^{2i\theta_i},e^{-2i\theta_i};q)_\infty\over \prod_{j=0}^s\prod_{\epsilon_{1,2}=\pm 1}(\alpha_{k+1}\alpha^{-1}_ke^{i\epsilon_1\theta_{k+1}+i\epsilon_2\theta_k};
q)_\infty}\\
H_{\bf n}=\prod_{i=1}^s{\Big({\alpha_{k+1}^2\over\alpha_0^2}q^{N_{k-1}+N_k-1};q\Big)_{n_k}\Big({\alpha_{k+1}^2\over\alpha_0^2}q^{2N_k};q\Big)_\infty
\over \Big(q^{n_{k+1}},{\alpha_{k}^2\over\alpha_0^2}q^{N_{k-1}+N_k};{\alpha_{k+1}^2\over\alpha_k^2}q^{n_k};q\Big)_\infty}\times
\prod_{\epsilon=\pm 1} \Big(\alpha_{s+1}\alpha_{s+2}^\epsilon q^{N_s},\alpha_{s+1}\alpha_{s+2}^\epsilon\alpha_0^{-2} q^{N_s};q\Big)_\infty^{-1}
\ee

\paragraph{Multivariable Racah polynomials.}
Similarly, at a special value of parameter, one can obtain discrete orthogonal polynomials, which are multivariable Racah polynomials \cite{Trat,GR2,Il}. Indeed,
consider
\be
{\alpha_{s+2}\over\alpha_{s+1}}=q^N
\ee
with a non-negative integer $N$ and put
\be
q^{J_k}=e^{i\theta_k}\alpha_k^{-1}
\ee
with some integer $J_k$'s. Then, the polynomials
\be
\mathfrak{R}_{\bf n}\left({\bf x},\Big|\,N,\alpha_i\right)
:= \prod_{k=1}^s\mathfrak{\tilde R}_{n_k}
\left(J_k-N_{k-1}\,\Big|\,{\alpha_k^2\over\alpha_0^2}q^{2N_{k_1}-1},\
{\alpha_{k+1}^2\over q\alpha_k^2},\ \alpha_k^2q^{J_{k+1}+N_{k-1}},\ J_{k+1}-N_{k-1}\right)\\
\mathfrak{\tilde R}_n(j|a,b,c,N):=(aq,bcq,q^{-N};q)_n\ q^{Nn/2}c^{-n/2}\ \mathfrak{R}_n (j;a,b,c,N;q)
\ee
are orthogonal on $0:=J_0\le J_1\le\ldots\le J_s\le J_{s+1}:=N$ with the discrete weight
\be
\mu_{\bf J}=\prod_{k=0}^s{\Big({\alpha_{k+1}^2\over\alpha_k^2};q\Big)_{J_{k+1}-J_k}\over (q;q)_{J_{k+1}-J_k}}{\Big(\alpha^2_{k+1};q\Big)_{J_{k+1}+J_k}\over
(q\alpha_k^2;q)_{J_{k+1}+J_k}}\Big(1-\alpha_k^2q^{2J_k}\Big)\left({\alpha_{k-1}\over\alpha_k}\right)^{2J_k}
\ee
and the norm squared
\be
||\mathfrak{R}_{\bf n}||^2={\Big(q\alpha_s^2,{\alpha_{s+1}^2\over\alpha_0^2};q\Big)_N\over (\alpha_{s+1}^2,q\alpha_0^2;q)_N}\
\Big({\alpha_{s+1}^2\over\alpha_0^2}q^N,\alpha_{s+1}^2q^N,{q^{-N}\over\alpha_0^2},q^{-N};q\Big)_{N_s}\
\left({\alpha_0^2\over\alpha_s^2}\right)^N\times\\
\times \prod_{k=1}^s {(q,\alpha_{k+1}^2;q)_{n_k}\Big({\alpha_s^2\over\alpha_0^2};q\Big)_{N_k+N_{k-1}}\over
\Big({\alpha_{k+1}^2\over q\alpha_0^2};q\Big)_{N_k+N_{k-1}}}{q\alpha_0^2-\alpha_{k+1}^2\over q\alpha_0^2-q^{2N_k}\alpha_{k+1}^2}
\ee

\subsection{Askey-Wilson functions}

The Askey-Wilson function is defined by \cite{R,KS,S,G}
\be\label{AWf1}
\Psi_\lambda (z|a,b,c,d):=\widehat h\Big(z\Big|{qa\lambda\over\tilde d}\Big)\ {({q/(\tilde d\lambda)},ab,ac,{qa/d};q)_\infty\over (\tilde a\tilde b\tilde c\lambda ;q)_\infty}\cdot \phantom{.}_8W_7\Big({\tilde a\tilde b\tilde c\lambda\over q};az,{a\over z},\tilde a\lambda,\tilde b\lambda,\tilde c\lambda\Big|q,{q\over\tilde d\lambda}\Big)
\ee
where
\be
\widehat h(z|a):=h(z+z^{-1}|a)
\ee
and $h(z|a)$ is defined in (\ref{h}).
Here we follow the definitions of $\tilde a, \tilde b, \tilde c, \tilde d$ as in eq.(\ref{6.9}).
This function is symmetric in $a$, $b$, $c$ and $q/d$.
Using Baily's transformation \cite{GaRa}, one can express the Askey-Wilson function through the sum of two balanced $_4\phi_3$ series,
\be\label{AWf2}
\Psi_\lambda (z|a,b,c,d)=\widehat h\Big(z\Big|{q\over d}\Big)\widehat h\Big(\lambda\Big|{q\over \tilde d}\Big)\ {(ab,ac;q)_\infty\over (q/(ad);q)_\infty}
\phantom{.}_4\phi_3\left(\begin{array}{c}
az,a/z,\tilde a\lambda,\tilde a/\lambda\\ ab,ac,ad\end{array}\Big|z=q\right)+\\
+\widehat h(z|a)\widehat h(\lambda|\tilde a)\ {(qb/d,qc/d;q)_\infty\over (ad/q;q)_\infty}\phantom{.}_4\phi_3\left(\begin{array}{c}
qz/d,q/(dz),q\lambda/\tilde d,q/(\lambda\tilde d)\\ qb/d,qc/d,q^2/(ad)\end{array}\Big|z=q\right)
\ee
We will sometimes also consider the Askey-Wilson function which differs by a factor
\be\label{AWf3}
\tilde\Psi_\lambda (z|a,b,c,d)={1\over \widehat h\Big(z\Big|\displaystyle{q\over d}\Big)}\Psi_\lambda (z|a,b,c,d)
\ee
which satisfies the same Askey-Wilson difference equation (\ref{AWdeq}) and, at $\lambda=\tilde a q^m$ with positive integer $m$, becomes the balanced terminating series $_4\phi_3$ and coincides with the Askey-Wilson polynomial up to a constant:
\be
\tilde\Psi_{\tilde a q^m} (z|a,b,c,d)=(-d)^mq^{m(m-1)/2}\,(abq^m,acq^m,bcq^m;q)_\infty\, P^{AW}_m\Big(z+{1\over z}\Big|a,b,c,d\Big)
\ee
It also has a single integral representation that can be read off from (\ref{intf}) and is a counterpart of the Mellin-Barnes representation (sec.5.2.2)
\be\label{AWfir}
\tilde\Psi_{\lambda} (z|a,b,c,d)={(a/z,b/z;q)_\infty\over (q/(cz),q/(dz);q)_\infty}{(q,ab\lambda/\tilde a,ac\lambda/\tilde a,bc\lambda/\tilde a)_\infty\over
(q\lambda/\tilde a)_\infty}\times\\
\times\int_{S^1} {\theta\Big({c\over y}\sqrt{ab\over q\lambda};q\Big)\Big(\sqrt{\lambda\over dc}{q\over zy},\sqrt{qab\lambda}zy,
\sqrt{c\over d\lambda}qy;q\Big)_\infty\over \Big(\sqrt{qb\lambda\over a}y,\sqrt{qa\lambda\over b}y,\sqrt{c\lambda\over d}{1\over y},\sqrt{ab\over q\lambda}
{1\over zy},{qyz\over\sqrt{dc\lambda}};q\Big)_\infty}{dy\over y}
\ee
where the integral runs over the unit circle $S^1$ and the renormalized Jacobi $\theta$-function is
\be
\theta(x;q):=(x,q/x;q)_\infty
\ee
Note that the Askey-Wilson function (\ref{AWf1}) satisfies the duality relation similar to (\ref{duality})
\be
\Psi_\lambda (z|a,b,c,d)=\Psi_z (\lambda|\tilde a,\tilde b,\tilde c,\tilde d)
\ee
and another duality relation
\be
\Psi_\lambda (z|a\xi,by,b/y,q\xi/a)=\Psi_\xi (y|a\lambda,bz,b/z,q\lambda/a)
\ee
Introduce a discrete set $\Gamma$ parameterized by an arbitrary parameter $\nu>0$: $\{\nu q^{-k}>1,\ k\in\mathbb{Z}\}$ and its dual $\tilde\Gamma$ parameterized by the dual parameter $\tilde\nu=\tilde d/(a\nu)$. Then, an essential property of the Askey-Wilson functions is that they are orthogonal for $\lambda\in\tilde\Gamma$,
\be\label{AWfo}
\Big<\Psi_\lambda,\Psi_{\lambda'}\Big>={1\over 2 i}\int_{S^1} \Psi_\lambda(z)\overline{\Psi_{\lambda'}(z)}{h(z^2|1)\over \widehat h(z|a,b,c,q/d)\theta(\nu z;q)\theta(\nu/z;q)}
{dz\over z}+2\pi\sum_{x\in\Gamma}\Psi_\lambda(x)\overline{\Psi_{\lambda'}(x)}W_k=\\
=||\Psi_\lambda||^2\delta_{\lambda,\lambda'},\ \ \ \ \ \ \ \ \lambda,\lambda'\in\tilde\Gamma
\ee
where
\be
W_k:={1\over \widehat h(\nu|a,b,c,q/d) (q,q;q)_{\infty}}{(1-q^{2k}\nu ^{-2})(a/\nu ,b/\nu ,c/\nu ,q/(d\nu );q)_k\over (q/(a\nu ),q/(b\nu ),q/(c\nu ),d/\nu ;q)_k}\ q^{k(k+1)}\Big({abc\nu ^2q\over d}\Big)^{-2k}
\ee
This gives the Askey-Wilson transform from functions with the scalar product (\ref{AWfo}) to the scalar product with the dual parameters $(a,b,c,d,\nu)\longrightarrow
(\tilde a,\tilde b,\tilde c,\tilde d,\tilde\nu)$:
\be
\Big<\Psi_\lambda(x),f(x)\Big>\longrightarrow \tilde f(\lambda)
\ee
This transform is unitary with inverse given by the Askey-Wilson functions with the dual parameters.

\subsection{Askey-Wilson functions at $|q|=1$\label{AWfq=1}}

Extending the notion of the Askey-Wilson function to the case of $|q|=1$ (for extension of the Askey-Wilson polynomials to the roots of unity, see \cite{SZ}), i.e. to $q=e^{2\pi i\tau}$ with a real parameter $\tau$ is not immediate, since the main building block, the Pocchammer symbol $(a;q)_\infty$ requires some regularization. The most convenient to regularize is, however, a differently normalized
quantity\footnote{The $q$-$\Gamma$-function is usually defined as \cite{GaRa}
\be
\Gamma_q(x):=(1-q)^{1-x}{(q;q)_\infty\over (q^x;q)_\infty}
\ee
Hence, the notation.} in logarithmic variables $x\to q^x$:
\be
{q^{-{(2x-1)^2\over 16}}\over (-q^x;q)_\infty}\longrightarrow \Gamma_\tau (x):=e^{i\gamma_\tau (x)},\ \ \ \ \ \ \ \gamma_\tau(x):={1\over 2i}\int_0^\infty {dy\over y}\left({2x-1\over y}-{\sinh \Big[(2x-1)\tau y\Big]\over
\sinh y\cdot\sinh\tau y}\right)
\ee
This function satisfies
\be\label{Gammar}
\Gamma_\tau(x+1)=2\cos\pi x\tau\cdot\Gamma_\tau(x)
\ee
and has the modular properties
\be\label{mod}
\Gamma_\tau(x)=\Gamma_{\tau^{-1}}(-\tau x+\tau/2+1/2),\ \ \ \ \ \ \ \ \Gamma_\tau(-x+1)=\Gamma^{-1}_\tau(x)
\ee
Now one can define \cite{S} the Askey-Wilson function at $|q|=1$ by the integral like (\ref{AWfir})
\be
\Psi_\lambda^\tau(x|a,b,c,d):=
{2\
\Gamma_\tau \Big({1\over 2\tau}-\gamma-{x\over 2}+1\Big)
\Gamma_\tau \Big({1\over 2\tau}-\delta-{x\over 2}+1\Big)\over
\Gamma_\tau \Big({1\over 2\tau}+\alpha-{x\over 2}\Big)
\Gamma_\tau \Big({1\over 2\tau}+\beta-{x\over 2}\Big)}\
\int_{\cal C} dz\ F_1(z)F_2(z)
\ee
where
\be
F_1(z):={\Gamma_\tau \Big({1\over 2\tau}-l+{x\over 2}+z-\delta-\gamma+1\Big)
\Gamma_\tau \Big({1\over 2\tau}-l-{x\over 2}-z+\alpha+\beta-1\Big)\over
\Gamma_\tau \Big(-{1\over 2\tau}+l-{x\over 2}-z-\delta-\gamma+1\Big)
\Gamma_\tau \Big(-{1\over 2\tau}+l+{x\over 2}+z+\alpha+\beta\Big)}\\
F_2(z):={\Gamma_\tau \Big(-{1\over 2\tau}+l-z+\gamma-\delta\Big)
\Gamma_\tau \Big(-{1\over 2\tau}+l+z+\alpha-\beta+{1\over 2}\Big)\over
\Gamma_\tau \Big(-{1\over 2\tau}-l+z+\gamma-\delta+1\Big)
\Gamma_\tau \Big(-{1\over 2\tau}-l-z+\alpha-\beta+{1\over 2}\Big)}
\ee
and
\be
q=e^{2\pi i\tau},\ \ \ \lambda=q^{2l},\ \ \ (a,b,c,d)=(q^{2\alpha},q^{2\beta},q^{2\gamma},q^{2\delta})
\ee
The contour ${\cal C}$ here is a deformation of the imaginary axis such that $F_1(z)$ and $F_2(z)$ are analytic on the stirp around ${\cal C}$ of width not less than 4. This function solves the Askey-Wilson difference equation (\ref{AWdeq}), and is a counterpart of the Mellin-Barnes representation (sec.5.2.2) at $|q|=1$.

\section{Orthogonal polynomials, matrix models and integrable systems}
\setcounter{equation}{0}
\subsection{Orthogonal polynomials and matrix models}

Orthogonal polynomials are a popular tool for investigation of the
eigenvalue matrix models \cite{BIPZ,GMMMO,KMMOZ,UFN3}.
Partition functions of these models are
\be\label{Hmm}
Z_N[d\mu;{\cal C}] = \Big<\!\!\Big<1\Big>\!\!\Big>_N := {1\over N!}\int_{\cal C} \Delta^2(z) \prod_{i=1}^N d\mu(z_i) =
\det_{1\leq i,j\leq N} \int z^{i+j-2} d\mu(z),
\ee
where $\Delta(z):=\det_{i,j} z_i^{j-1}=\prod_{i<j}(z_i-z_j)$ is the Vandermonde determinant. Hereafter, for the sake of brevity, we omit the integration contour ${\cal C}$ from the arguments of the partition function in most cases.
By suitable rearrangement of the Vandermonde determinant, we can show that these partition functions
are consecutive product of the norms of monic polynomials
orthogonal with respect to the measure $d\mu(x)$,
\be
\int {\cal P}_n(x) {\cal P}_m(x)\, d\mu(x) = ||{\cal P}_n||^2 \cdot \delta_{m,n},
\ee
\be
Z_N[d\mu]=\prod_{i=0}^{N-1} ||{\cal P}_i||^2,
\ee
In addition, the monic orthogonal polynomials can be understood as expectation value of the resolvent determinant with respect to the matrix integral (see (\ref{Pmm})),
\be
{\cal P}_n(x) = \frac{\Big<\!\!\Big<\prod_{i=1}^n(x-z_i)\Big>\!\!\Big>_n}{\Big<\!\!\Big<1\Big>\!\!\Big>_n}
= x^n + \ldots .
\ee

Since multiplication by $x$ is a Hermitian operation in the Hilbert space expanded in terms of orthogonal polynomials:
\be
\int {\cal P}_n(x) \cdot x {\cal P}_m(x) d\mu(x) = 0 \ \ {\rm for} \ \ m<n-1,
\ee
these polynomials satisfy the three-term recursion relation,
\be\label{3termmm}
x{\cal P}_n(x) = {\cal P}_{n+1}(x) + K_n {\cal P}_n(x) + R_n {\cal P}_{n-1}(x).
\ee
From
\be
\int {\cal P}_{n}(x)\cdot x{\cal P}_{n-1}(x) \, d\mu(x) = ||{\cal P}_n||^2 = R_n\cdot ||{\cal P}_{n-1}||^2,
\ee
we deduce that
\be
||{\cal P}_n||^2 = ||{\cal P}_0||^2 \prod_{k=1}^n R_k, \ \ \mbox{and} \ \
Z_N[d\mu]= ||{\cal P}_0||^{2N} \prod_{n=1}^{N-1} \prod_{k=1}^{n} R_k.
\ee

To make a $q$-deformation of the eigenvalue matrix model, we only need to change the integral into the Jackson sum
\be
\int_0^1 f(z) dz \ \longrightarrow \
\int f(z) d_q(z) = \sum_{k=0}^\infty (1-q) q^k f(q^k),
\ee
and the Vandermonde determinant is $q$-deformed when matrix models are promoted to the $\beta$-ensembles,
with $\Delta^{2\beta}(z) \longrightarrow \Delta_{\beta}^{(2)}(z;q)$ equal to
\be
\Delta_{\beta}^{(2)}(z;q):=\prod_{i<j}\Big(z_i/z_j;q\Big)_\beta \Big(z_j/z_i;q\Big)_\beta
\ee
for positive integer $\beta \in N$, and to
\be
\Delta_{\beta}^{(2)}(z;q):=\prod_{i<j}{\Big(z_i/z_j;q\Big)_\infty\over \Big(q^\beta z_i/z_j;q\Big)_\infty}{\Big(z_j/z_i;q\Big)_\infty\over \Big(q^\beta z_j/z_i;q\Big)_\infty}
\ee
otherwise.

\subsection{Orthogonal polynomials and integrable systems}

The relation between matrix models and integrable systems allows one to make a bridge between orthogonal polynomials and integrable systems. Essentially, the system of orthogonal polynomials is nothing but the Baker-Akhiezer function (up to an exponential factor) of an integrable forced hierarchy of the Toda type (i.e. with a discrete type variable) \cite{GMMMO,KMMOZ,versus}. Let us explain how it works in the simplest case of the model (\ref{Hmm}).

First of all, one can notice \cite{GMMMO} that the matrix model partition function (\ref{Hmm}) is the $\tau$-function of the forced \cite{Kaup,KMMOZ,versus} Toda chain hierarchy. Indeed, this $\tau$-function has the generic T\"opliz form \cite{versus}
\be\label{tau}
\tau_n=\det_{1\le i,j\le n}\partial^{i+j-2}C(t)
\ee
where $C(t)$ is the moment matrix that depends on infinitely many time variables $t_k$, and $\partial$ denotes the derivative w.r.t. the first time, the dependence on higher times being determined from the compatibility condition
\be\label{C}
{\partial C\over\partial t_k}=\partial^k C
\ee
and the initial conditions are
\be
\tau_0=1,\ \ \ \ \ \ \ \ \ \tau_{-1}=0
\ee
This $\tau$-function satisfies an infinite set of Hirota bilinear identities: the differential-difference bilinear Toda equation
\be
\tau_n\partial^2\tau_n-(\partial\tau_n)^2=\tau_{n+1}\tau_{n-1}
\ee
and an infinite hierarchy of differential bilinear equations of the KP type w.r.t. to the time variables $t_k$ \cite{versus}.

The general solution to (\ref{C}) in the Fourier representation is
\be\label{Ct}
C(t)=\int_{\cal C}d\mu_0(z)\exp\Big(\sum_k t_kz^k\Big)
\ee
which can be easily related to the general point of the infinite-dimensional Grassmannian describing the space of solutions to this infinite hierarchy \cite[Appendix ???]{versus}. On the other hand, comparing (\ref{Hmm}) with (\ref{tau}) and (\ref{Ct}), one immediately arrives at an identification of the orthogonality measure $d\mu(z)$ with the integrand in (\ref{Ct}):
\be
d\mu(z)=d\mu_0(z)\exp\Big(\sum_k t_kz^k\Big)
\ee
and the Toda chain $\tau$-function (\ref{tau}) with the matrix model partition function (\ref{Hmm}):
\be
\tau_n=Z_n\Big[d\mu(z)=d\mu_0(z)\exp\Big(\sum_k t_kz^k\Big)\Big]
\ee
i.e. the integrable flows are just a proper parametrization of the orthogonality measures so that the concrete solution of the integrable hierarchy (i.e. the concrete point of the Grassmannian) is fixed by the choice of $d\mu_0(z)$. In matrix model theory, this choice is usually done via a set of Virasoro constraints (in fact, the lowest constraint plus integrable equations are sufficient to fix the whole algebra of constraints; this lowest constraint is called the string equation).

There is, however, much more direct connection between the orthogonal polynomials and the integrable systems, not just through the measure, or through the $\tau$-function being a product of the norms: the orthogonal polynomials are nothing but (up to a factor) the Baker-Akhiezer functions of the integrable system \cite{GMMMO}. Indeed, the Lax representation for the Toda system is
\be
{\partial {\rm {\bf L}}\over\partial t_k}={1\over 2}\Big[{\cal R}{\rm {\bf L}}^k,{\rm {\bf L}}\Big]
\ee
where the Lax matrix is
\be
{\rm {\bf L}}_{mn}:=\delta_{m,n-1}+K_n\delta_{m,n}+R_{n+1}\delta_{m,n+1}
\ee
and the ${\cal R}$-matrix acts on the matrix $A_{ij}$ in accordance with
\be
{\cal R}A_{ij}:=\left\{
\begin{array}{cl}
A_{ij}&\hbox{if }i>j\\
-A_{ij}&\hbox{if }i\le j
\end{array}
\right.
\ee
Then, the 3-term relation (\ref{3termmm}) can be treated as an eigenvalue problem
\be\label{Laxe}
{\rm {\bf L}}\Psi(z;t)=z\Psi(z;t)
\ee
where $z$ plays the role of the spectral parameter. The eigenfunction of the Lax operator is called the Baker-Akhiezer function, and it is just the orthogonal polynomial in our case. The only subtlety is that in (\ref{Laxe}) the normalization of the eigenfunction is not fixed, and the normalization of the Baker-Akhiezer function has an additional factor of $\exp\Big({1\over 2}\sum_k t_kz^k\Big)$ as compared with the orthogonal polynomial:
\be
\Psi_n(z;t)=\exp\Big({1\over 2}\sum_k t_kz^k\Big){\cal P}_n(z;t)
\ee

\subsection{Orthogonal polynomials as a bridge to matrix models of Kontsevich type}

Representations of matrix model partition functions of type (\ref{Hmm}) in terms of orthogonal polynomials allows one to establish their connections with matrix models of the Kontsevich type. Let us consider again $\tau_n=Z_n\Big[d\mu(z)=d\mu_0(z)\exp\Big(\sum_k t_kz^k\Big)\Big]$ with the sum including the term $t_0\log z$ and make the change of time variables $t_k$ to the Miwa variables $\lambda_a$, $a=1,...,N$ and $N$ is large enough \cite{Miwa,GKM}:
\be\label{Miwa1}
t_k={1\over k}\sum_a\lambda_a^{-k},\ \ \ \ \ \ \ \ t_0=\sum_a \log \lambda_a
\ee
Now, using the integral representation in formula (\ref{Hmm}), one can easily derive \cite{versus,AMMC} that
\be\label{key}
Z_n\Big[d\mu(z)=d\mu_0(z)\exp\Big(\sum_k t_kz^k\Big)\Big]=\tau_n[d\mu(z)]=\tau_n[d\mu_0(z)]\times {\det_{a,b}{\cal P}_{n+a-1}(\lambda_b)\over\Delta(\lambda)}
\ee
where ${\cal P}_n$ here are the monic orthogonal polynomials w.r.t the measure $d\mu_0(z)$.

This is the key formula that we need. Now our strategy will be to reduce a matrix model to the eigenvalue model of type (\ref{Hmm}), and a matrix model of the Kontsevich type to the r.h.s. of (\ref{key}). In fact, one can consider more general models than (\ref{Hmm}), this question will be addressed elsewhere.

\subsubsection{Hermitian versus Kontsevich models}

Let us consider now as our reference example the partition function of the Gaussian Hermitian one-matrix model \cite{UFN3,MMchar}
\be\label{H1mm}
{\bf Z}^{\mathbb{H}}_n(t):={1\over \hbox{Vol}(U_n)}\int D_nM \exp\Big(-{1\over 2}\tr M^2+\sum_{k=0}t_k\tr M^k\Big)=Z_n\Big[\exp\Big(-{1\over 2}z^2+\sum_k t_kz^k\Big);\mathbb{R}\Big]
\ee
i.e. the Toda chain $\tau$-function (\ref{tau}) with $d\mu_0(z)=\exp\Big(-z^2/2\Big)$ and the integration contour ${\cal C}=\mathbb{R}$. Here $\hbox{Vol}(U_n)$ is the Haar measure of the unitary group $U(n)$ and $D_n M$ is the Cartesian  measure on the $n\times n$ Hermitian matrices. This formula implies that one has to consider the Hermite polynomials in (\ref{key}) in order to deal with this model. Note that the monic Hermite polynomials have the integral representation
(see (\ref{irHer}))
\be\label{mH}
{\cal H}_n(x):=He_n(x)=2^{-n/2}H_n\Big({x\over\sqrt{2}}\Big)={i^n\over\sqrt{2\pi}}\int_{-\infty}^{+\infty}dyy^n e^{x^2/2-ixy-y^2/2}
\ee
Note also that
\be
Z_n\Big[\exp\Big(-{1\over 2}z^2\Big);\mathbb{R}_+\Big]=(2\pi)^{n/2}\prod_{k=0}^{n-1}k!:={\bf Z}^{\mathbb{H}}_n(0)
\ee
The Kontsevich type matrix model is given by an integral over the $N\times N$ Hermitian matrix $X$ and depends on an external $N\times N$ matrix $\Lambda$
\be\label{KH}
{\bf Z}^{K\mathbb{H}}_N(n,\Lambda):=(2\pi)^{(n-N)/2}i^{nN}\Big(\prod_{k=0}^{n-1}k!\Big)\times{\displaystyle{e^{\Tr \Lambda^2\over 2}}\over\hbox{Vol}(U_N)} \int D_NX\exp\Big(-{1\over 2}\Tr X^2+n\Tr\log X-i\Tr X\Lambda\Big).
\ee
Using the Itzykson-Zuber formula \cite{IZ},
\be
\int D_NU\exp\Big(\Tr XUYU^\dagger\Big)=\hbox{Vol}(U_N){\det_{a,b} e^{x_ay_b}\over\Delta(x)\Delta(y)},
\ee
where $D_NU$ is the integral over the unitary $N\times N$ matrix $U$, and $x_a$, $y_a$ are eigenvalues of the matrices $X$, $Y$,
we find that the partition function depends only on the eigenvalues $\lambda_a$ of the matrix $\Lambda$. The explicit result is given as
\be\label{KH1}
{\bf Z}^{K\mathbb{H}}_N(n,\Lambda)={\bf Z}^{\mathbb{H}}_n(0)\times{\det_{a,b}\Phi_a^{\mathbb{H}}(\lambda_b)\over\Delta(\lambda)},
\ee
where
\be\label{PhiH}
\Phi_a^{\mathbb{H}}(z):={i^n\over \sqrt{2\pi}}e^{z^2\over 2}\int_{-\infty}^{+\infty}dx x^{a-1}e^{-x^2/2+n\log x-ixz}={\cal H}_{n+a-1}(z),
\ee
see (\ref{mH}).
Now, using (\ref{key})-(\ref{H1mm}) and (\ref{KH1})-(\ref{PhiH}), we obtain that
\be\boxed{
{\bf Z}^{\mathbb{H}}_n(t)={\bf Z}^{K\mathbb{H}}_N(n,\Lambda)},
\ee
with
\be\label{Miwa2}
t_k={1\over k}\Tr \Lambda^{-k},\ \ \ \ \ \ \ \ t_0=\Tr \log \Lambda.
\ee

\subsubsection{Complex Gaussian versus rectangular Kontsevich models}

Another example of a similar correspondence is the complex matrix model case. Let us consider instead of (\ref{H1mm}) the complex Gaussian one-matrix model \cite{MMC,MMchar,IMM2}
\be\label{Cmm}
{\bf Z}^{\mathbb{C}}_n(\nu,t):={1\over \Big(\hbox{Vol}(U_n)\Big)^2}\int D_n{\cal M}D_n{\cal M}^\dagger \exp\Big(-\tr ({\cal M}{\cal M}^\dagger)+{\nu}\tr\log ({\cal M}{\cal M}^\dagger)+\sum_{k=0}t_k\tr ({\cal M}{\cal M}^\dagger)^k\Big)=\\
=Z_n\Big[z^{\nu}\exp\Big(-z+\sum_k t_kz^k\Big);\mathbb{R}_+\Big]
\ee
where the integral runs over the complex matrix ${\cal M}$ with the Cartesian integration measure.
This formula implies that one has to consider this time the Laguerre polynomials in (\ref{key}). The monic Laguerre polynomials have the integral representation (see (\ref{irLag}))
\be\label{mL}
{\cal L}_n^{(\nu)}(x)= (-1)^n e^x x^{-\nu/2}\int_0^\infty dz e^{-z}z^{\nu/2+n}J_\nu\Big(2\sqrt{xz}\Big)
\ee
Note that
\be
Z_n\Big[z^{\nu}\exp(-z);\mathbb{R}_+\Big]=\prod_k^{n-1}(k+\nu)! \ k!:={\bf Z}^{\mathbb{C}}_n(\nu,0)
\ee

Now we compare (\ref{Cmm}) with the corresponding rectangular complex Kontsevich model \cite{AMMC}
\be\label{KC}
{\bf Z}^{K\mathbb{C}}_{N_1,N_2}(n,\Xi\Xi^\dagger)={\mathfrak{N}_{N_1,N_2}\ e^{-\Tr\!_{N_1} (\Xi\Xi^\dagger)}\over \hbox{Vol}(U_{N_1})\hbox{Vol}(U_{N_2})} \int D{\cal X}D{\cal X}^\dagger e^{\Big(-\Tr\!_{N_1} ({\cal X}{\cal X}^\dagger)
+\Tr\!_{N_1} (\Xi{\cal X}^\dagger)+\Tr\!_{N_1} ({\cal X}\Xi^\dagger)+(n+\nu)\Tr\!_{N_1}\log (\Xi\Xi^\dagger)\Big)}\\
\mathfrak{N}_{N_1,N_2}=2^{N_1(N_1+1)-\nu(\nu-1)/2}\left( \prod_l^{n-1}(l+\nu)! \ l! \right)
\left(\prod_{k=0}^{\nu-1}k!\right)\ \ \ \ \ \ \ \ \ \ \ \ \ \ \ \ \nu:=N_1-N_2
\ee
where the integral runs over the complex rectangular $N_1\times N_2$ matrix ${\cal X}$ and its $N_2\times N_1$ Hermitian conjugate ${\cal X}^\dagger$. Hence, the external matrix $\Xi$ is this case an $N_1\times N_2$ matrix. We could also choose all the traces in (\ref{KC}) to be taken over the matrices of size $N_2\times N_2$ by inverting the order of all terms in the exponentials in (\ref{KC}). Without loss of generality, we put $N_1\ge N_2$. The matrix $\Xi$ can be rotated by unitary matrices to a matrix with $N_2$ non-zero entries $\xi_a$ on a diagonal, and $\Tr\!_{N_1}f(\Xi\Xi^\dagger)=\sum_{a=1}^{N_2}f(\xi_a\xi^\dagger_a)$.

This integral can be also dealt with using the generalization of the Itzykson-Zuber formula \cite{IZg}
\be
\int DU_1\int DU_2 \exp\Big({1\over 2}\Tr \Big[U_1{\cal X}U_2{\cal Y}+{\cal Y}^\dagger U_2^\dagger{\cal X}^\dagger U_1^\dagger \Big]\Big)=
\hbox{Vol}(U_{N_1})\hbox{Vol}(U_{N_2}){2^{N_1(N_2-1)-\nu(\nu+1)/2}\over\prod_{k=0}^{\nu-1}k!\prod_{a}^{N_2} (x_ay_a)^\nu}{\det_{a,b}J_\nu(x_ay_b)\over
\Delta(x^2)\Delta(y^2)}
\ee
where $U_1$ and $U_2$ are the unitary matrices of sizes $N_1\times N_1$ and $N_2\times N_2$ respectively and $x_a$, $y_a$ are the eigenvalues of the matrices ${\cal X}$ and ${\cal Y}$ respectively.
Hence, one immediately obtains
\be\label{KC1}
{\bf Z}^{K\mathbb{C}}_{N_1,N_2}(n,\Xi,\Xi^\dagger)={\bf Z}^{\mathbb{C}}_n(\nu,0)\times {\det_{a,b\le N_2}\Phi_a^{\mathbb{C}}(\xi_b\xi_b^\dagger)\over\Delta(\xi\xi^\dagger)},
\ee
where
\be\label{PhiC}
\Phi_a^{\mathbb{C}}(z)=e^{z}z^{-\nu/2}\int_{0}^{+\infty}dx x^{a-1-\nu/2}e^{-x+(n+\nu)\log x}J_\nu(2\sqrt{xz})=(-1)^{n+a-1}{\cal L}^{(\nu)}_{n+a-1}(z),
\ee
see (\ref{mL}). Now, using (\ref{key}), (\ref{Cmm}) and (\ref{KC1})-(\ref{PhiC}), we obtain that
\be\boxed{
{\bf Z}^{\mathbb{C}}_n(\nu,t)={\bf Z}^{K\mathbb{C}}_{N_1,N_2}(n,\Xi\Xi^\dagger)},
\ee
with
\be\label{Miwa2}
t_k={1\over k}\Tr\!_{N_1} (\Xi\Xi^\dagger)^{-k},\ \ \ \ \ \ \ \ t_0=\Tr\!_{N_1} \log (\Xi\Xi^\dagger).
\ee

\section{Orthogonal polynomials, $6j$-symbols and knots}
\setcounter{equation}{0}
\subsection{Colouring knots with orthogonal polynomials}

The old classical problem of effective constructing knot polynomials \cite{knotpols,Con,HOMFLYPT} has got a new momentum during last decades and now is one of the most challenging problems in modern mathematical physics.

One of the most effective ways to construct knot polynomials is within the Reshetikhin-Turaev approach (RT) that was suggested in \cite{RT,Wit} and further developed in \cite{inds,braids}. The idea is to consider a plane projection of a given knot, to put an $R$-matrix or its inverse at each crossing depending on the type of the crossing and take a specifically defined trace of the product of the $R$-matrices (in fact, this is the usual trace with insertions of specific operators). This trace gives a knot polynomial (the HOMFLY-PT polynomial for the $SL_q(N)$ $R$-matrices). The topological invariance of this expression is guaranteed by invariance w.r.t. to three Reidemeister moves \cite{Reid}. The second Reidemeister move is satisfied, since the opposite crossing is described by the inverse $R$-matrix. The third move is nothing but the Yang-Baxter equation for the $R$-matrices. The most non-trivial is the first move, it is guaranteed by a properly defined trace.

In fact, technically there are two different descriptions: every knot has a representation as a braid so that there has been developed a procedure of calculating the knot polynomial for its representation as a braid \cite{braids}. There is also a large class of arborescent knots \cite{Con,Cau}, which can be calculated in an easier way \cite{arbor,MMSqracah}. Moreover, there is also a mixture of these two approaches \cite{fingers} that allows one to evaluate knot polynomials in the most effective way known so far.

The general feature of all these approaches is manifest by constructing the $R$-matrix acting on the tensor product $V_1\otimes V_2$ in the basis of irreducible representations emerging in the decomposition of this product: $V_1\otimes V_2\longrightarrow\oplus_i W_i$. In fact, since some of the representation can appear a few times, one works in the space of intertwining operators: $V_1\otimes V_2=\oplus_W {\cal M}^{V_1V_2}_{W} W$. In this basis, some of the $R$-matrices can be chosen to be diagonal, then, the remaining ones are given by conjugation of these diagonal matrices with the Racah matrices, or the Wigner $6j$-symbols, i.e. matrices intertwining the maps
\be
(V_1\otimes V_2)\otimes V_3  \longrightarrow V_4 \qquad \text{and} \qquad V_1\otimes (V_2\otimes V_3) \longrightarrow V_4
\ee
The Racah matrices \cite{Racah} are always orthogonal (or unitary in the complex case) and, hence, satisfy some orthogonality relations. This means that one can colour the knot with orthogonal polynomials. However, one has to check each time that the Reidemeister moves are satisfied, which is quite a non-trivial constraint. For the Racah matrices, they are guaranteed. Hence, we consider in this section various examples of the Racah matrices that give rise to various orthogonal systems and to knot polynomials. Note that, depending on the concrete case, the Racah matrices can be discrete orthogonal polynomials (hence, the discrete index), or instead orthogonal functions (hence, continuous index and distributions). The reference example of the first case is given by the Racah polynomials, while the second example consists of two cases: the Askey-Wilson functions, associated with unitary infinite-dimensional representations at $0<q<1$, and the Racah ``matrices" constructed from the Barnes functions, corresponding to the case of $|q|=1$.

\paragraph{3-strand braid.} To complete this subsection, we briefly describe the Racah matrices that emerge in description of knot polynomials.
First consider as an example the knot presented by a three-strand braid. Then, its HOMFLY polynomial is described by the formula \cite{braids}
\be
H_V(q,A)=\Tr_q\Big(R_1^aR_2^b\ldots\Big)=\sum_{Q\vdash 3|V|} \Big(R_1^{(Q)}\Big)^a_{i_1i_2}\Big(R_2^{(Q)}\Big)^b_{i_2i_3}\ldots \Big(R_*^{(Q)}\Big)^*_{i_ni_1}
\text{dim}_q\,Q,\ \ \ \ \ A:=q^N
\ee
where $Q\in V^{\otimes 3}$. The $R$-matrices are equal to ($W\in V^{\otimes 2}$)
\be\label{R1R2}
\Big(R_1^{(Q)}\Big)_{ij}=\pm q^{\kappa_{W_i}}\delta_{ij},\ \ \ \ \ \Big(R_2^{(Q)}\Big)_{ij}=U(V,V,V;Q)_{ik}\Big(R_1^{(Q)}\Big)_{kj}U(V,V,V;Q)_{ij}
\ee
Here $U(V,V,V;Q)_{ik}$ are the Racah matrices for $V^{\otimes 3}\to Q$,  $\kappa_{W}$ is the value of the second Casimir operator in representation $W\in V^{\otimes 2}$, the sign of the eigenvalue of the $R$-matrix depends on whether $W$ lies in the symmetric or antisymmetric square of $V$, and $\text{dim}_q\,Q$ is the quantum dimension of representation $Q$.

\paragraph{Arborescent knots.}
Arborescent knots forms a distinguished class of knots introduced by J.Conway in 1967 \cite{Con,Cau}. His purpose was to give a new classification of knots, which could be more systematic and logic than already existed. {\it Arborescent  knot} is a closure of arborescent tangle, which can be obtained from a trivial tangles with the help of certain operations. Any arborescent knot can be represented as a weighted tree with any valent vertices \cite{arbor}. Edges of the tree are 4-strand braids with two strands having an opposite orientation to the other two. In the vertices of the tree, each edge connects with adjacent edges with 2 strands, therefore, every edge connects with two other edges.

Thus, to deal with the arborescent knots, one needs 4-strand braids, where two strands have an opposite orientation to the other two. From the point of view of representation theory, this means that there is a representation $V$ associated with one orientation and $\bar V$ (a conjugate representation) with the opposite one. Closure of these braids into an arborescent knot corresponds to a projection on a singlet representation $\varnothing$, which always lives in $V^{\otimes2}\otimes\bar V^{\otimes2}$.

Now we have three R-matrices corresponding to the generators of the braid group: $R_{1}$, $R_{2}$, and $R_{3}$. Usually one only has to require that $R_1$ and $R_3$ are commuting: $[R_1,R_3]=0$, i.e. that they have a common system of eigenfunctions. However, the condition of projecting onto the singlet implies that, in the case of multiplicity-free representations (in particular, symmetric ones), they just coincide: $R_{1}$ = $R_{3}$ \cite{arbor}. At the same time, for $R_1$ and $R_2$ (\ref{R1R2}) is still correct, only now $W$ can lie in $V\otimes\bar V$ as well.

Thus, considering the case of arborescent knots or links with components colored by same representations, whatever be the representation $V$, one has only two different $R$-matrices $R_{1}$: $T$ acting in  $V\otimes V$, and $\bar T$ acting in $V\otimes \bar V$. Accordingly, there are also two Racah matrices, because all other Racah matrices arising in the arborescent calculus can be reduced to these two cases:
\be\label{Smat}
\bar S := U(V,\bar V,V; V)=\left\{U_{ij}\left[\begin{array}{cc} V& \bar V \\ V& \bar V \end{array}\right]  \right\}\, ,\qquad
S := U(V,V, \bar V; V)=\left\{U_{ij}\left[\begin{array}{cc} V& V \\ \bar V& \bar V \end{array}\right]  \right\}\,.
\ee

Given a Young diagram corresponding to $V$, the Young diagram corresponding to $\bar V$ is a completion to $V$ of the rectangular diagram of height $N$ and width $\lambda_1$, where $V=\{\lambda_1\geq\lambda_2\geq\ldots\}$. These two Racah matrices are unitary and are related via
\be
\bar S = \bar T^{-1} S T^{-1} S^\dagger \bar T^{-1}
\label{SvsbS}
\ee

\bigskip

The elements of (quantum) Racah matrix are related with the (quantum) 6j-symbols by a simple normalization:
\be
U_{ij} = \epsilon_{\{V\}} \sqrt{\text{dim}_q\,V_{i} \ \text{dim}_q\,V_{j}} \cdot \left\{
\begin{array}{ccc}
	V_1& V_2&V_{i} \\
	V_3& V_4&V_{j}
\end{array}
\right\},
\ee
where $\text{dim}_q\,V$ is a quantum dimension of representation $V$ and $\epsilon_{\{V\}}$ is a sign.

\subsection{Racah polynomials and symmetric representations}

Thus, as we already stressed there are many orthogonal systems but it is not simple to distinguish those that provide the Yang-Baxter equations, which we needs for the purposes of knot theory. In order to find such orthogonal systems, one needs to use the group theory: the orthogonal systems coming as $6j$ symbols, or Racah matrices do provide orthogonal systems with the required properties. First, we consider here the finite-dimensional representations of $SL_q(2)$ when the orthogonal system is just a system of discrete orthogonal polynomials, the Racah polynomials considered in section 6.

\subsubsection{Racah for $SL_q(2)$}

In the case of $SL_q(2)$ representations, all 6j-symbols, or Racah matrix were computed by A.Kirillov and N.Reshetikhin in \cite{KR} (see also \cite{Racah1,AG}). The answers can be given in terms of the $q$-hypergeometric function $_4\phi_3$, that is, in terms of the $q$-Racah polynomials (\ref{Racahs}), hence, the name of these latter. Let us introduce a new notation for the $q$-Racah polynomial, which will be more convenient throughout this section:
\be\label{Rachat}
\widehat{\mathfrak{R}}_n(x;a,b,c,d) := {_4\phi_3}\left(\begin{array}{c}q^{-2n},q^{2a+2b+2n+2},q^{-2x},q^{2x+2c+2d+2}\\
	q^{2a+2},q^{2c+2},q^{2b+2d+2} \end{array}\Big|\,q^2,q^2\right)
\ee

\be
\hspace{-5mm}
\left\{
\begin{array}{ccc}
	r_1& r_2&i \\
	r_3&r_4&j
\end{array}
\right\} \sim \widehat{\mathfrak{R}}_{\frac{r_1+r_3}{2}-j}{\footnotesize
	\left(\frac{r_1+r_3}{2}-i \ ; \ {-}r_3{-}1,{-}r_2{-}1,{-}\frac{1}{2}(r_1{+}r_2{+}r_3{+}r_4{+}2),\frac{1}{2}(r_2{-}r_1{+}r_4{-}r_3)
	\ \right)}
\nonumber
\ee

The matrix $T$ is a diagonal matrix, see (\ref{R1R2}):
\be
T = {\rm diag}\left( \epsilon_{R_m}\, q^{c_2(X_m)} \right), \ [r_1]\otimes [r_2] = \oplus_{m} [r_1+r_2-2m] =: \oplus_{m}X_m,
\ee
where $c_2(R)$ is an eigenvalue of the second Casimir operator and $\epsilon_{R}$ is a sign.

The matrix $S$ coincides with $\bar S$ in the case of $SL_q(2)$, since $r=\bar r$ in this case, and so does $T$ coinciding with $\bar T$. The Racah matrix is equal to
\be
S_{ij}	\ = \  (-1)^{r+i+j}\dfrac{\sqrt{(2i+1)(2j+1)}\,[r]!^2[2r+1]!}{[r{-}i]![r{-}j]![r{+}i{+}1]![r{+}j{+}1]!} \cdot
\widehat{\mathfrak{R}}_{r-j}\left(r-i\,;\, {-}r{-}1, {-}r{-}1, {-}2r{-}2, 0\,\right)
\label{Ssl2}
\ee

\subsubsection{Racah for symmetric representation of $SL_q(N)$}

For symmetric representations of $SL_q(N)$, the diagonal matrices $T$ and $\bar T$, which become different, take the following form
\be
T = {\rm diag}\left((-1)^{m+1}\dfrac{q^{-r^2+m^2+m}}{A^r}\right), \ m=0...r, \qquad
\bar T = {\rm diag}\Big((-q^{m-1}A)^m\Big), \ m=0...r
\ee
where $A:=q^N$.

$\bullet$ Let us consider the matrix $\bar S$ in the case of $V_1=V_3=[r]$, $V_2=V_4=[\bar r]$ so that $V_{12}$ and $V_{23}$ are representations of the type $\mathcal{V}_n=[2n,n^{N-2}]$ of $SL_q(N)$, which emerge in the decomposition:
\be
[r]\otimes[\bar r]=\oplus_{n=0}^r\mathcal{V}_n
\ee
We denote $V_{12}=\mathcal{V}_i\to i$ and $V_{23}=\mathcal{V}_j\to j$, and the Racah matrix is symmetric in $i$ and $j$.
In this case, the explicit formula was given in \cite{Racah2} and \cite{MMSpret} in terms of partial sums. In the paper \cite{MMSaracah}, it was shown that the corresponding 6j-symbols can be written as
\be
\label{f3}
\left\{
\begin{array}{ccc}
	r&\bar r&i \\
	r&\bar r&j
\end{array}
\right\}
= \dfrac{[i]!^2[j]!^2[2r{+}N{-}1]![N{-}1]![N{-}2]!}
{[r{+}i{+}N{-}1]![r{+}j{+}N{-}1]![r{-}i]![r{-}j]![i{+}j{-}r]![i{+}j{-}r{+}N{-}2]!}
\cdot \nonumber  \\ \cdot {_4\Phi_3}
\left[
\begin{array}{c}
	i{-}r,i{-}r,j{-}r,j{-}r \\
	1{-}2r{-}N,i{+}j{-}r{+}1,i{+}j{-}r{+}N{-}1
\end{array}\Bigg| \, q, q^2
\right],
\ee
where we have used a $q$-hypergeometric function of exponential parameters defined as follows
\be
{}_r\Phi_s\left(\begin{array}{c} a_1,\ldots, a_r \\ b_1,\ldots,b_s \end{array}\Bigg|\,q,\,z\right)  =   {}_r\phi_s\left(\begin{array}{c} q^{2a_1},\ldots, q^{2a_r} \\ q^{2b_1},\ldots, q^{2b_s} \end{array}\Bigg|\,q^2,\,z\right).
\ee

In order to obtain alternative expressions for the 6j-symbols, one can use Sears' transformations for terminating balanced $_4\Phi_3\left[\ldots;q,q^2\right]$ (see \cite{GaRa,KV}):
\be
\label{transf}
{_4\Phi_3}
{
	\left[
	\begin{array}{c}
		x,y,z,n \\
		u,v,w
	\end{array}\Bigg| \, q, q^2
	\right]
}    =  \dfrac{[v{-}z{-}n{-}1]![u{-}z{-}n{-}1]![v{-}1]![u{-}1]!}{[v{-}z{-}1]![v{-}n{-}1]![u{-}z{-}1]![u{-}n{-}1]!} \cdot {_4\Phi_3}
{
	\left[
	\begin{array}{c}
		w-x,w-y,z,n \\
		1-u+z+n,1-v+z+n,w
	\end{array}\Bigg| \, q, q^2
	\right]
}
\ee
where the numbers $x,y,z,u,v,w$ are integer, and the balanced series condition
\be
x+y+z+n+1 = u+v+w
\label{balance}
\ee
is fulfilled. Also one can use the invariance of $_4\Phi_3\left[\ldots;q^2,q^2\right]$ under permutations of $x,y,z,n$ or $u,v,w$. Then, one gets the following expression for the $6j$-symbols:
\be
\!\!\!\!\!\!\!
\left\{
\begin{array}{ccc}
	r&\bar r&i \\
	r&\bar r&j
\end{array}
\right\}
=
\dfrac{[i]![j]![N{-}1]![N{-}2]!}{[i{+}N{-}2]![j{+}N{-}2]!} \dfrac{[r]![r+N-2]![2r+N-1]!}{[r{-}i]![r{-}j]![r{+}i{+}N{-}1]![r{+}j{+}N{-}1]!}
\cdot   {_4\Phi_3}
{\tiny
	\left[
	\begin{array}{c}
		j{-}r,{-}r{-}j{-}N{+}1,i{-}r,{-}r{-}i{-}N{+}1 \\
		-r,{-}r{-}N{+}2,{-}2r{-}N{+}1
	\end{array}\Big| \,  q, q^2
	\right]
}
\label{hg3}
\ee
By expanding $_4\Phi_3\left[\ldots;q^2,q^2\right]$ in terms of q-factorials, one gets a generalization of the most commonly encountered formulas in the literature about $SL(2)$ $6j$-symbols. Such expansions are related by Sears' transformations (\ref{transf}) of the balanced hypergeometric series. In terms of the $q$-Racah polynomials, formula (\ref{hg3}) takes the form

{\footnotesize
	\be
	\hspace{-6mm}
	\left\{
	\begin{array}{ccc}
		r&\bar r&i \\
		r&\bar r&j
	\end{array}
	\right\}
	\ = \ \dfrac{[i]![j]![N{-}1]![N{-}2]!}{[i{+}N{-}2]![j{+}N{-}2]!} \dfrac{[r]![r+N-2]![2r+N-1]!}{[r{-}i]![r{-}j]![r{+}i{+}N{-}1]![r{+}j{+}N{-}1]!} \cdot
	\widehat{\mathfrak{R}}_{r-j}\left(r-i\,;\, {-}r{-}1, 1{-}r{-}N, {-}2r{-}N, 0\,\right)
	\label{qrac}
	\ee
}
It was also found a three-term relation for these $6j$-symbols
\be
\label{3rel}
a_1 \left\{
\begin{array}{ccc}
	r&\bar r&i \\
	r&\bar r&j-1
\end{array}
\right\} + a_2 \left\{
\begin{array}{ccc}
	r&\bar r&i \\
	r&\bar r&j
\end{array}
\right\} + a_3 \left\{
\begin{array}{ccc}
	r&\bar r&i \\
	r&\bar r&j+1
\end{array}
\right\} = 0
\ee
\be
a_1 = [j]^2\,[j-r-1]\,[r+j+N-1]\,[N+2j] \nonumber \\
a_3 = [j-r]\,[j+N-1]^2\,[N+r+j]\,[N+2j-2] \nonumber \\
a_1+a_2+a_3 = -[i]\,[i+N-1]\,[N+2j-2]\,[N+2j-1]\,[N+2j].
\ee

\bigskip

\bigskip

$\bullet$ Now let us consider the matrix $S$, when $V_1=V_2=[r]$, $V_3=V_4=[\bar r]$
so that $V_{23}$ is still of the type $\mathcal{V}_n=[2n,n^{N-2}]$ since belongs to the decomposition of $[r]\otimes [\bar r]$, while $V_{12}$ belongs to
\be
[r]\otimes[r]=\oplus_{n=0}^r[r+n,r-n]
\ee
We denote $V_{12}=[r+i,r-i]\to i$ and $V_{23}=\mathcal{V}_j\to j$ and the Racah matrix is no longer symmetric in $i$ and $j$.
Below we assume that $i<j$, otherwise one needs to change $i\longleftrightarrow j$ in all answers. In this case, the manifest formulas also can be found in  \cite{Racah2} and \cite{MMSpret}. With the help of hypergeometric notations and Sears' transformation (\ref{transf}), one can convert this into
\be
\label{hg4}
\left\{
\begin{array}{ccc}
	r& r&i \\
	\bar r&\bar r&j
\end{array}
\right\} =
\dfrac{[r]!^2[N{-}1]![N{-}2]![2r+N-1]!}{[r{-}i]![r{-}j{+}N{-}2]![r{+}i{+}N{-}1]![r{+}j{+}N{-}1]!} \cdot {_4\Phi_3}
\left[
\begin{array}{c}
	j{-}r,{-}r{-}j{-}1,i{-}r,{-}r{-}i{-}N{+}1 \\
	-r,{-}r,{-}2r{-}N{+}1
\end{array}\Bigg| \, q, q^2
\right],
\ee
which, in terms of the q-Racah polynomials (\ref{Rachat}), takes the form
{\footnotesize
	\be
	\label{qr4}
	\!\!\!\!\!\!\!\!\!
	\left\{
	\begin{array}{ccc}
		r& r&i \\
		\bar r&\bar r&j
	\end{array}
	\right\} =
	\dfrac{[r]!^2[N{-}1]![N{-}2]![2r+N-1]!}{[r{-}i]![r{-}j{+}N{-}2]![r{+}i{+}N{-}1]![r{+}j{+}N{-}1]!} \cdot \widehat{\mathfrak{R}}_{r-j}\left(r-i\,;\, {-}r{-}1, {-}r{-}1, {-}2r{-}N, 0\,\right)
	\ee}
The corresponding three-term recurrence relation is
\be
\label{3rel}
b_1 \left\{
\begin{array}{ccc}
	r& r&i \\
	\bar r&\bar r&j-1
\end{array}
\right\} - b_2 \left\{
\begin{array}{ccc}
	r& r&i \\
	\bar r&\bar r&j
\end{array}
\right\} + b_3 \left\{
\begin{array}{ccc}
	r& r&i \\
	\bar r&\bar r&j+1
\end{array}
\right\} = 0
\ee
with
\be
b_1 = [j]^2\,[j+r+1]\,[r-j+N-1]\,[2j+2] \nonumber \\
b_3 = [r-j]\,[j+1]^2\,[N+r+j]\,[2j] \nonumber \\
b_1+b_2+b_3 = [i]\,[i+N-1]\,[2j]\,[2j+1]\,[2j+2].
\ee

\subsection{Askey-Wilson functions, infinite-dimensional representations and Hikami invariants}

As we already mentioned, an advantage of obtaining orthogonal systems for constructing knot polynomials via the Racah matrices is that it guarantees the third Reidemeister move, i.e. the Yang-Baxter equation. In this way, one can construct both orthogonal polynomials and orthogonal functions. The former case is described by the Racah matrices of finite-dimensional representations and was discussed in the previous subsection. The other case corresponds to the Racah matrices of infinite-dimensional representations (of non-compact groups), and, in that case, one has to check convergency of the proper infinite sums or integrals. One meets this kind of problems when constructing the knot invariants from affine algebras (see, however, \cite{AKMMMOZ}). However, in the case of finite-dimensional groups, the integrals can be defined. We do not enter the details and discuss here only the corresponding Racah matrices, postponing explanations of how to construct from them knot invariants to a separate publication.

\subsubsection{Infinite-dimensional unitary representations of $SU_q(1,1)$}

We start from the case of $0<q<1$ and discuss the Racah matrices for unitary infinite-dimensional irreducible representations of $SU_q(1,1)$ \cite{G}. In this case, there are five different series of representations \cite{BK,KV}: positive and negative discrete series $\pi^{\pm}_k$, principal $\pi^P_{\rho,\epsilon}$, complementary $\pi^C_{\lambda,\epsilon}$ and strange series, where $k>0$, $0\le\epsilon<1$, $0\le\rho\le -{\pi\over 2\log q}$, and $-{1\over 2}<\lambda<-\epsilon$ ( $-{1\over 2}<\lambda<\epsilon-1$) for $0\le\epsilon <{1\over 2}$ (${1\over 2}\le\epsilon\le 1$). For the sake of illustrative purposes, we write down the Racah matrices in three different cases \cite{G}. To this end, we need to know how the tensor products of the representations decompose into irreducible irreps in these cases. These decompositions are:
\be\label{dc1}
\pi^+_{k_1}\otimes\pi^+_{k_2}=\oplus_{j=0}^\infty\ \pi^+_{k_1+k_2+j}
\ee
\be\label{dc2}
\pi^+_{k_1}\otimes\pi^-_{k_2}=\int_0^{-{\pi\over 2\log q}}d\rho\ \pi^P_{\rho,\{k_1-k_2\}}\ \ \ \ \ \ \ \ \ \hbox{for }k_1+k_2\ge{1\over 2},\ \ k_1-k_2\ge -{1\over 2}
\ee
\be\label{dc3}
\pi^P_{\rho,\epsilon}\otimes\pi^-_k=\int_0^{-{\pi\over 2\log q}}d\sigma\ \pi^P_{\sigma,\{\epsilon -k\}}\ \oplus_j\ \pi^-_{j-\epsilon+k}
\ee
\be\label{dc4}
\pi_k^+\otimes\pi^P_{\rho,\epsilon}=\int_0^{-{\pi\over 2\log q}}d\sigma\ \pi^P_{\sigma,\{\epsilon +k\}}\ \oplus_j\ \pi^+_{j+\epsilon+k}
\ee
Here the second subscript $\{\epsilon\}$ in $\pi^P_{\rho,\{\epsilon\}}$ is the fractional part of $\epsilon$. Below, we use hat in order to denote the procedure of replacing $q$ with $q^2$ and using the logarithmic parameters as in (\ref{Rachat}): $\widehat f(q^x|a):=f(q^{2x}|q^{2a})$. The Racah matrices corresponding to described products (\ref{dc1})-(\ref{dc4}) are:
\begin{itemize}
\item for the product $\pi^+_{k_1}\otimes\pi^+_{k_2}\otimes\pi^-_{k_3}$, when the representations emerging in the decomposition of the first two factors, (\ref{dc1}) are labelled by non-negative integers $j$ and those emerging in the decomposition of the second two factors, (\ref{dc2}) are labelled by the continuous parameter $\rho$,
\be
U_{j\rho}\left[\begin{array}{cc} k_1& k_2 \\ k_3& \tau \end{array}\right]
=\mathfrak{N}_1\cdot \widehat P^{AW}_j\Big(q^{i\rho}+q^{-i\rho}\Big|k_2-k_3+{1\over 2},k_2+k_3-{1\over 2},k_1+i\tau,k_1-i\tau\Big)\\
\mathfrak{N}_1:={2\pi q^{-2jk_1}q^{j(j-1)/2}(q^{2k_1+2k_2+2k_3-1+2j +2i\tau},q^{2k_1+2k_2+2k_3-1+2j -2i\tau};q^2)_\infty\over
(q^{4k_1};q^2)_j(q^2,q^2,q^{4k_3},q^{4k_1+4k_2+4j},q^{4i\tau},q^{-4i\tau};q^2)_\infty}\times\\
\times{\theta(q^{2k_1+2k_2-2k_3+1+2j +2i\tau};q^2)
\theta(q^{2k_1+2k_2-2k_3+1+2j -2i\tau};q^2)\over\sqrt{(q^{4k_1+4k_2+2j},q^{1-2k_1-2k_2+2k_3-2j+2i\tau},q^{1-2k_1-2k_2+2k_3-2j-2i\tau};q^2)_j}}
\ee
\item for the product $\pi^+_{k_1}\otimes\pi^P_{\rho,\epsilon}\otimes\pi^-_{k_3}$, when the representations emerging in the decompositions of the both first two factors, (\ref{dc4}) and the second two factors, (\ref{dc3}) are labelled by the continuous parameters $\sigma$, $\zeta$,
\be
U_{\sigma\zeta}\left[\begin{array}{cc} k_1& (\rho,\epsilon) \\ k_3& \tau \end{array}\right]
=\mathfrak{N}_2\cdot \widehat\Psi_{q^{i\zeta}}\Big(q^{i\sigma}\Big|k_1+i\rho,k_3+i\tau,k_3-i\tau,1-k_1+i\rho\Big)\\
\mathfrak{N}_2:={2\pi\theta(q^{1+2k_1-2k_3+2\epsilon+2i\tau};q^2)\theta(q^{1+2k_1-2k_3+2\epsilon-2i\tau};q^2)\over
(q^2,q^{4k_1},q^{4k_3},q^{4i\tau},q^{-4i\tau},q^{1+2\epsilon+2k_1+2i\sigma},q^{1+2\epsilon+2k_1-2i\sigma},
q^{1-2\epsilon+2k_3+2i\zeta},q^{1-2\epsilon+2k_3-2i\zeta};q^2)_\infty}
\ee
\item for the product $\pi^+_{k_1}\otimes\pi^+_{k_2}\otimes\pi^P_{\rho,\epsilon}$, when the representations emerging in the decomposition of the first two factors, (\ref{dc1}) are labelled by non-negative integers $j$ and those emerging in the decomposition of the second two factors, (\ref{dc4}) are labelled by the continuous parameter $\sigma$,
\be
U_{j\sigma}\left[\begin{array}{cc} k_1& k_2 \\ (\rho,\epsilon)& \tau \end{array}\right]
=\mathfrak{N}_3\cdot \widehat P^{AW}_j\Big(q^{i\sigma}+q^{-i\sigma}\Big|k_2+i\rho,k_2-i\rho,k_1+i\tau,k_1-i\tau\Big)\\
\mathfrak{N}_3:=2\pi q^{{1\over 2}j(j-4k_1-1)}\ {\prod_{\varepsilon=\pm} (q^{1-2k_1-2k_2-2\epsilon+2i\varepsilon\tau};q^2)_\infty
\prod_{\varepsilon_{1,2}=\pm} (q^{2k_1+2k_2+2j+2i\varepsilon_1 \rho+2i\varepsilon_2 \tau};q^2)_\infty\over
(q^{4k_1};q^2)_j(q^2,q^{4k_1+4k_2+4j},q^{1-2\epsilon+2i\rho},q^{1-2\epsilon-2i\rho},q^{4i\tau},q^{-4i\tau};q^2)_\infty}\times\\
\times\sqrt{(q^{1-2k_1-2k_2-2\epsilon-2j+2i\tau},q^{1-2k_1-2k_2-2\epsilon-2j-2i\tau};q^2)_j\over (q^{4k_1+4k_2+2j};q^2)_j}
\ee
\end{itemize}
One can use these Racah matrices in order
to construct knot polynomials. We will return to this problem elsewhere.

\subsubsection{$|q|=1$ and Hikami invariants}

As we observed in s.7.3, there is an extension of the Askey-Wilson functions, i.e. the functions that solve the Askey-Wilson difference equation (\ref{AWdeq}) to
the case of $|q|=1$, i.e. in the case of real $\tau$, $q=e^{2\pi i\tau}$. In fact, their modular properties (that follow from (\ref{mod})) provide that the Askey-Wilson functions in this case also satisfy the other, modular inverted equation with $\tilde q:=e^{2\pi i/\tau}$. This is a consequence of the duality  with respect to interchanging $q\leftrightarrow\tilde q$, i.e. $\tau\to 1/\tau$ \cite{Fad,FKV,FK,PT} and is associated with the modular double of $SL_q(2)$ (which was introduced by L. Faddeev \cite{F}), or with the quantum algebra $SL_q(2)\times SL_{\tilde q}(2)$ (see also \cite{KhLS}).
In this case, a corresponding class of one-parametric representations with weights $\alpha_i^\vee$ was described in \cite{PT,BT,T01} (see also \cite{DFad}), their Racah matrices \cite{PT} being expressed through the Askey-Wilson functions at $|q|=1$ (see also \cite{IDS} for the non-deformed case):
\be\label{PT}
U_{\alpha_s^\vee\alpha_t^\vee}\left[\begin{array}{cc} \alpha_1^\vee& \alpha_2^\vee \\ \alpha_3^\vee& \alpha_4^\vee \end{array}\right]
\sim
\Psi^{\tau}_{q^{\alpha_t}}\Big(-2\alpha_s\Big| q^{\alpha_4-\alpha_3+Q/b}, q^{\alpha_1-\alpha_2+Q/b}, q^{\alpha_3+\alpha_4},q^{\alpha_1+\alpha_2}\Big)\\
\hbox{where }\ \ \ \ \ \ Q=b+b^{-1},\ \ \ \ \ \ \tau=b^2,\ \ \ \ \ \alpha_i:={\alpha_i^\vee\over b}-{Q\over 2b}
\ee
Here $\alpha_s^\vee$ and $\alpha_t^\vee$ denote the intermediate representations in the Racah matrices.

Having constructed the Racah matrices, one can further find the $R$-matrix \cite{F,Kash2,BT} (see also \cite{Kash,Hikn}) and further knot invariants \cite{GMM}. These knot invariants are known and called Hikami invariants, or integral state model \cite{Hik,Hikn,Z,AK}. For the simplest knots they look like
\be\label{Htr}
\mathbb{H}(3_1,x)=e^{{3x^2\over \hbar}+{i\pi x\over \hbar}}
\ee
\be\label{Hfe}
\mathbb{H}(4_1,x)=\int dy \frac{\Phi_b(x-y)}{\Phi_b(y)}e^{2\pi i x(2y-x)}
\ee
 \be
      \mathbb{H}(5_2,x)=\int dz \frac{e^{\pi i (z^2-x^2)}}{\Phi_b(z)\Phi_b(z-x)\Phi_b(x+z)}
    \ee
where Faddeev's quantum dilogarithm \cite{Fad3} is defined as
\be
\Phi_b(x):={\Big(-e^{2\pi bx}\sqrt{q};q\Big)_\infty\over\Big(-e^{2\pi x/b}\sqrt{\tilde q};\tilde q\Big)_\infty}
\ee
Note that the Racah matrix (\ref{PT}) is nothing but the modular kernel that gives rise to the modular transformation of the conformal blocks \cite{PT} in the two-dimensional conformal theory with central charge $c=1+6Q^2$, the parameter $\alpha^\vee$ of representations being associated with the conformal dimensions of the fields in the conformal block in accordance with $\Delta=\alpha^\vee(Q-\alpha^\vee)$.

Being the Askey-Wilson function at $|q|=1$, the Racah matrix (\ref{PT}) satisfies the Askey-Wilson difference equation (\ref{AWdeq}) (see s.\ref{AWfq=1}). This equation determines the modular kernel as a function of intermediate dimensions ($\alpha_s^\vee$ and $\alpha_t^\vee$) and coincides with the equation for the modular kernel obtained within the $pq$-duality approach \cite{GMMpq}, where the conformal blocks are treated as eigenfunctions of the monodromy check operators \cite{check}.

\section{Conclusion}

In this paper, we have reviewed properties of the orthogonal polynomials, mostly hypergeometric polynomials, and discussed several extensions in application to various branches of mathematical physics: to integrable theories, matrix models, two-dimensional conformal theories and knot theory. Our discussion was inevitably short and sketchy, since we tried to include all these topics into a general framework. However, the approaches described here can be immediately pushed forward.

In particular, as was explained in sec.9, with the knowledge of Racah matrices, one can construct knot invariants. The knot invariants made from the Racah matrices for finite-dimensional representations of $SL_q(2)$ are known, these are the Jones polynomials. However, only recently it was discovered \cite{Racah2,MMSpret,MMSqracah} how to construct the knot invariants colored with symmetric representations of $SL_q(N)$, these are the corresponding HOMFLY-PT polynomials \cite{HOMFLYPT}. The problem of constructing the Racah matrices in non-symmetric representations, both rectangular (when there are no multiplicities) \cite{rect} and non-rectangular \cite{nonrect} (see also \cite{knotebook}) as well as their interpretation in terms of hypergeometric polynomials remains unsolved.
Similarly, the knot polynomials colored with representations of the modular double of $SL_q(2)$ are also known, these are Hikami invariants \cite{Hik}. However, the extension of these invariants to other representations (see, e.g., \cite{F1}) or to groups of higher ranks has not been done yet. It can be immediately done using results of sec.9.3.2 and the trick that allowed us to come from $SL_q(2)$ Racah matrices to $SL_q(N)$ ones in sec.9.2. Similarly, one could construct knot invariants from infinite-dimensional
unitary representations of $SU_q(1,1)$ and higher rank groups using the results of sec.9.3.1. We shall return to these issues elsewhere.

Another issue related to the same Racah matrices for the modular double is related to their interpretation as modular kernels of the modular transformations of conformal blocks. This also waits for its extension to higher groups, which corresponds to special conformal blocks of the $W$-algebras. It would be also interesting to interpret these results within the $pq$-duality approach \cite{GMMpq}, where the conformal blocks are treated as eigenfunctions of the monodromy check operators \cite{check}.

Note that absolutely mysterious here remain extensions of these modular kernels to conformal blocks of the $q$-Virasoro algebra. This issue has to be explained within the more general context of the Ding-Iohara-Miki algebras \cite{DIM,DIM1} and the corresponding network matrix models \cite{DIM1,MMZ}.

Note that the integral representations that we discussed in s.5, and the related modular transformations of secs.5 and 9 have a lot to do with the new emerged topic of constructing solutions to the Painl\'eve and discrete Painl\'eve equations in terms of the conformal blocks of the Virasoro and $q$-Virasoro algebras. This problem also will be addressed in a separate publication \cite{MMZ}.

At last, there is a rather traditional question of relating the eigenvalue type matrix models with those of the Kontsevich type. So far, this correspondence included at most rectangular complex matrix models and has been established by using the Laguerre polynomials (see sec.8.3). The correspondence inspired by more general orthogonal polynomials is not known yet, but it probably could be constructed. Another direction is to construct a generalization of these relations to the $q$-deformed case even in the case of the $q$-Laguerre polynomials \cite{C}.

\section*{Acknowledgements}

Our work was partly supported by the grant of the Foundation for the Advancement of Theoretical Physics ``BASIS" (A.Mor. and A.S.), by RFBR grants 16-01-00291 (A.Mir.) and 16-02-01021 (A.Mor.), by joint grants 17-51-50051-YaF, 15-51-52031-NSC-a, 16-51-53034-GFEN, 16-51-45029-IND-a (A.M.'s and A.S.), by grant 16-31-60082-mol-a-dk (A.S.). C.T.Chan was supported by the grant of Ministry of Science and Technology of Taiwan
under the grants 104-2923-M-002-003-MY3 and 106-2112-M-029-003. The discussions with M. Ismail on the Shohat-Favard theorem is greatly acknowledged.

\appendix

\section{Hypergeometric series}
\def\theequation{A.\arabic{equation}}
\setcounter{equation}{0}
In this Appendix, we collect all the definitions and notation related to the ($q$-)hypergeometric series that we use throughout the paper. The main textbooks are \cite{BE,GaRa,BE2}.

\subsection{Generalized hypergeometric functions}

The generalized hypergeometric functions of one variable are usually defined as an analytical continuation (or resummations) of the generalized hypergeometric series,
\be\label{hyper}
\phantom._rF_s\left(\begin{array}{c} a_1,\ldots, a_r \\ b_1,\ldots,b_s \end{array}\Big|\,z\right)
:= \sum_{n=0}^\infty \frac{(a_1)_n\ldots (a_r)_n}{(b_1)_n \ldots (b_s)_n}\cdot\frac{z^n}{n!}
= \sum_{n=0}^\infty \dfrac{(a_1,\ldots ,a_{p})_n}{(b_1,\ldots ,b_q)_n}\frac{z^n}{n!} ,
\ee
with $(a)_n := \dfrac{\Gamma(a+n)}{\Gamma(a)}= \prod_{k=0}^{n-1}(a+k),\ \ \ \  (a_1,\ldots ,a_i)_n :=(a_1)_n\cdot\ldots \cdot(a_i)_n.$

\vspace{0.2cm}
One way to obtain an integral representation of the generalized hypergeometric function is to use the Euler integral formula,
\be
 _{r+1}F_{s+1} \left( \begin{array}{c} a_1,\ldots, a_r, a \\ b_1,\ldots,b_s, b \end{array}
\Big|\,z \right) \hspace{8cm} \\
= \dfrac{\Gamma(b)}{\Gamma(a) \Gamma(b-a)} \int_0^1 t^{a-1} (1-t)^{b-a-1} \
_{r}F_{s} \left(\begin{array}{c} a_1,\ldots, a_r \\ b_1,\ldots,b_s \end{array}\Big|\,tz\right) dt.
\ee
In the special case $r=s+1$, we get
\be
\label{intrep}
{}_{q+1}F_q \left(
\begin{array}{c} a_0,a_1,\ldots, a_{q} \\ b_1,\ldots,b_q \end{array}\Big|\,z
\right) \hspace{8cm} \\
= \dfrac{\Gamma(b_1) \Gamma(b_2) \ldots, \Gamma(b_q)}{\Gamma(a_1) \Gamma(a_2) \ldots, \Gamma(a_{q}) \cdot \Gamma(b_1 - a_1) \Gamma(b_2 - a_2)\ldots, \Gamma(b_q - a_{q})} \times
\\ \times \prod_{j=1}^q \left( \int_0^{t_{j+1}}  dt_j \,t_j^{a_{j}-a_{j-1}-1} (1 - t_j/t_{j+1})^{b_j-a_{j}-1}\right) t_1^{a_0} (1-t_1 z)^{-a_0}.
\ee
where $t_{q+1}:=1$.

There are also recursions which change the difference $r-s$, but they involve
exponential weights:
\be\label{ir2}
\phantom._{r+1}F_s\left(\begin{array}{c} a_1,\ldots, a_{r+1} \\ b_1,\ldots,b_s \end{array}\Big|\,z\right)={1\over\Gamma(a_{r+1})}\int_0^\infty dt\ e^{-t}t^{a_{r+1}-1}\,\phantom._{r}F_s\left(\begin{array}{c} a_1,\ldots, a_{r} \\ b_1,\ldots,b_s \end{array}\Big|\,zt\right)
\ee
and
\be
\phantom._rF_{s+1}\left(\begin{array}{c} a_1,\ldots, a_r \\ b_1,\ldots,b_{s+1} \end{array}\Big|\,z\right)={\Gamma(b_{s+1})\over 2\pi i}\int_{c-i\infty}^{c+i\infty} dt\ e^{t}t^{-b_{s+1}}\,
\phantom._{r}F_s\left(\begin{array}{c} a_1,\ldots, a_{r} \\ b_1,\ldots,b_s \end{array}\Big|\,{z\over t}\right)
\ee
where $c>0$. These formulas are proved by comparing the coefficients at the both sides of formulas using the integral representations of the $\Gamma$-functions
\be\label{Gamma}
\Gamma (z)=\int_0^\infty dt\ e^{-t}t^{z-1},\ \ \ \ \ \ \ \ \ \ {1\over \Gamma (z)}={1\over 2\pi i}\int_{c-i\infty}^{c+i\infty} dt\ e^{t}t^{-z}
\ee

\subsection{Basic hypergeometric series}

The $q$-analogue of the generalized hypergeometric series, called Heine's basic hypergeometric series, or hypergeometric $q-$series is defined as,
\be\label{qhg}
\phantom._r\phi_s\left(\begin{array}{c} a_1,\ldots, a_r \\ b_1,\ldots,b_s \end{array}\Big|\,q,\,z\right)
:= \sum_{k=0}^\infty \frac{(a_1,\ldots,a_r;q)_k}{(b_1,\ldots,b_s;q)_k}\cdot\frac{z^k}{(q;q)_k}\Big[(-1)^kq^{k(k-1)/2}\Big]^{1+s-r}
\ee
where
\be
(a_1,\ldots,a_r;q)_k:=\prod_{i=1}^r(a_i;q)_k,\ \ \ \ \ \ \ (a;q)_k:=\prod_{i=0}^{k-1}(1-aq^i),\ \ \ \ \ \ \ (a;q)_0=1,\ \ \ \ \ \ \ (a;q)_\infty:=\prod_{i=0}^{\infty}(1-aq^i)
\ee
Note that, at non-natural $k$, one usually substitutes $(a;q)_k$ with
\be
(a;q)_k\to {(a;q)_\infty\over (aq^k;q)_\infty}
\ee
We apply this trick in the paper in various places.

The basic hypergeometric series $_{r+1}\phi_r$ is called {\it balanced} if
\be
\prod_{i=1}^{r+1} a_i=q\prod_{j=1}^rb_j
\ee
and {\it very-well-poised} if $a_2=q\sqrt{a_1}$, $a_3=-q\sqrt{a_1}$, $b_1=\sqrt{a_1}$, $b_2=-\sqrt{a_1}$, $b_j=qa_1/a_{j+1}$ for $j>2$.
We denote the very-well-poised hypergeometric series
\be\label{W}
{_{r+1}W_r}(a_1;a_4,a_5,\ldots,a_{r+1}|q,z)
:=_{r+1}\phi_r\left(\begin{array}{c}a_1,q\sqrt{a_1},-q\sqrt{a_1},a_4,\ldots,a_{r+1}\\
\sqrt{a_1},-\sqrt{a_1},qa_1/a_4,\ldots,qa_1/a_{r+1}\end{array}\Big|\,q,z\right)
\ee

\section{Very classical orthogonal polynomials}
\def\theequation{B.\arabic{equation}}
\setcounter{equation}{0}

In this Appendix, we list some of the very classical polynomials: the Jacobi, Laguerre and Hermite polynomials, and their properties. These are the polynomials that we discuss in the paper.

\subsection{List of the very classical orthogonal polynomials}

\vspace{.5cm}

\subsubsection*{\framebox{Jacobi polynomials}}

\begin{itemize}
\item {\bf The Jacobi polynomial as a hypergeometric polynomial}
\be
\label{jac}
P_n^{(a,b)}(x) := \dfrac{(a+1)_n}{n!} \
{_2}F_1\left( \begin{array}{c} -n, \ \ a+b+n+1  \\  a+1  \end{array} \ \Big| \ \frac{1-x}{2} \right)=\\
=
P_n^{(a,b)}(1)\
{_2}F_1\left( \begin{array}{c} -n, \ \ a+b+n+1  \\  a+1  \end{array} \ \Big| \ \frac{1-x}{2} \right).
\ee
\item {\bf weight and orthogonal relation}
\be
\label{jacort}
\int_{-1}^{1}  \, P_m^{(a,b)}(x) \, P_n^{(a,b)}(x) \,(1-x)^a\,(1+x)^b \ dx = \hspace{6cm}\\
\dfrac{2^{a+b+1}}{2n+a+b+1}\dfrac{\Gamma(n+a+1)\,\Gamma(n+b+1)}{\Gamma(n+a+b+1)\, n!} \delta_{m,n}, \qquad a,b>-1
\ee
\item {\bf differential equation}
\be
(1-x^2) P_n^{(a,b)''} + [b-a-x(a+b+2)]P_n^{(a,b)'} + n(n+a+b+1) P_n^{(a,b)} = 0
\ee

\item{\bf 3-term relation}
\be
\dfrac{2n(n+a+b)(2n+a+b-2)}{2n+a+b-1} \, P_n^{(a,b)}(x) = \hspace{12cm}   \\ = \left[ (2n+a+b)(2n+a+b-2)x + a^2-b^2 \right] \, P_{n-1}^{(a,b)}(x) \hspace{11cm} \\
- \dfrac{2(n+a-1)(n+b-1)(2n+a+b)}{2n+a+b-1} P_{n-2}^{(a,b)}(x) \hspace{11cm}
\ee

\item {\bf Rodriguez formula}
\be
P_n^{a,b}(x)={(-1)^n\over 2^nn!}(1-x)^{-a}(1+x)^{-b}{d^n\over dx^n}\Big[(1-x)^{a+n}(1+x)^{b+n}\Big]
\ee

\item{\bf generating function}
\be
\dfrac{2^{a+b}}{R \ (1 - t + R)^a \ (1 + t + R)^b} = \sum_{n=0}^\infty P_n^{(a,b)}(x) \ t^n.\ \
R:= \sqrt{1 - 2 t x + t^2}.
\ee

\item {\bf specializations of the Jacobi polynomials}

\begin{itemize}
\item for $a=b=\nu-1/2$ we get {\bf Gegenbauer} polynomials,
\be
C_n^{(\nu)} (x) := \dfrac{(2 \nu)_n}{( \nu + 1/2)_n} \ P_n^{(\nu - 1/2, \nu - 1/2)} (x).
\ee
\vspace{0.cm}
\item For $a=b=0$ we get {\bf Legendre} polynomials, \\

\item for $a=b=\pm \frac{1}{2}$ we get {\bf Chebyshev} polynomials.
\vspace{0.cm}
\end{itemize}

\end{itemize}
\vspace{0.2cm}

\subsubsection*{\framebox{Laguerre polynomials}}

\begin{itemize}
\item {\bf Laguerre polynomial as a hypergeometric polynomial}
\be\label{Lag}
L_n^{(a)}(x) = \dfrac{(a+1)_n}{n!} \ {_1}F_1\left( \begin{array}{c} -n \\  a+1  \end{array} \Big| x \right)
\ee

\item {\bf weight and orthogonal relation}
\be
\int_{0}^{\infty}  \, L_m^{(a)}(x) \, L_n^{(a)}(x) \, x^a\, e^{-x} dx
= \dfrac{\Gamma(n+a+1)}{n!} \delta_{m,n}, \qquad a>-1
\ee

\item {\bf differential equation}
\be
x L_n^{(a)''} + (a+1-x)L_n^{(a)'} + n L_n^{(a)}  = 0
\ee

\item {\bf 3-term relation}
\be
L_n^{(a)}(x) = \left( 2 - \dfrac{x-a+1}{n} \right) \, L_{n-1}^{(a)}(x) - \left( 1 + \dfrac{a-1}{n} \right) L_{n-2}^{(a)}(x)
\ee

\item{\bf Rodriguez formula}
\be
L_n^{a}(x)={e^xx^{-a}\over n!}{d^n\over dx^n}\Big[e^{-x}x^{n+a}\Big]
\ee

\item{\bf generating function}
\be
\dfrac{1}{(1 - t)^{a+1}} \ e^{- \frac{x t}{1 -t}} =
\sum_{n=0}^\infty L_n^{(a)} (x) {t^n}.
\ee

\item {\bf as a limit of the Jacobi polynomial}
\be
\lim_{b\to \infty} P_n^{(a,b)}\left( 1-2b^{-1}x \right) = L_n^{(a)}(x)
\ee
\end{itemize}

\bigskip

\subsubsection*{\framebox{Hermite polynomials}}

\begin{itemize}
\item  {\bf Hermite polynomial as a hypergeometric polynomial}
\be\label{herm}
H_n(x) = (2x)^n \ {_2}F_0\left( \begin{array}{c} -n/2, \ -(n-1)/2  \\  -   \end{array} \Big| \, -\dfrac{1}{x^2}\right)=\left\{
\begin{array}{cl}
{(-1)^m n!\over m!} \phantom{.}{_1}F_1\left( \begin{array}{c} -m \\  {1\over 2}  \end{array} \Big| x^2 \right)&n=2m\\
{(-1)^m n!\over m!}\ 2x \phantom{.}{_1}F_1\left( \begin{array}{c} -m \\  {3\over 2}  \end{array} \Big| x^2 \right)&n=2m+1
\end{array}
\right..
\ee
We can check this by deriving the corresponding differential equation.

\item {\bf weight and orthogonal relation}
\be
\int_{-\infty}^{\infty} H_m(x) \, H_n(x) \, e^{-x^2} \, dx = 2^n\,n! \sqrt{\pi} \, \delta_{m,n}.
\ee

\item {\bf differential equation}
\be
H_n^{''} - 2x H_n^{'} + 2n H_n  = 0.
\ee

\item {\bf 3-term relation}
\be
H_{n+1}(x) -  2xH_n(x) + 2n H_{n-1}(x) = 0.
\ee

\item{\bf Rodriguez formula}
\be
H_n(x)=(-1)^ne^{x^2}{d^n\over dx^n}e^{-x^2}
\ee

\item{\bf generating function}
\be
e^{2 x t - t^2} = \sum_{n=0}^\infty H_n(x) \dfrac{t^n}{n!}.\
\ee

\item {\bf as a limit of Jacobi polynomial}
\be
\lim_{a\to \infty} a^{-\frac{n}{2}} \ P_n^{(a,a)}\left( a^{-\frac{1}{2}}x \right) = \dfrac{H_n(x)}{2^n\,n!}.
\ee
\end{itemize}

\subsection{Orthogonality of the Jacobi polynomials}

For illustrative purposes, we reproduce here a proof of the orthogonality of the Jacobi polynomials (\ref{jacort}).

Let us denote l.h.s. of (\ref{jacort}) by $J$, expand Jacobi polynomials as hypergeometric series (\ref{jac}) and get
\be
\label{jac1}
J = \dfrac{(a+1)_n(a+1)_m}{n!\,m!} \sum_{k,l=0}^{\infty} \dfrac{(-n)_k(a+b+n+1)_k(-m)_l(a+b+m+1)_l}{(a+1)_k(a+1)_l\,k!\,l!\,2^{k+l}} \times \\
\times \int_{-1}^{1} (1-x)^{k+l+a}\,(1+x)^b\,dx
\ee
To evaluate the integral we use Euler integral representation for beta function:
\be
\int_{-1}^{1} (1-x)^{k+l+a}\,(1+x)^b\, dx \  \stackrel{x=1-2t}{=} \ -\int_{1}^{0} 2^{k+l+a+b+1} \, t^{k+l+a}\,(1-t)^b\, dt \\
= 2^{k+l+a+b+1} \, \dfrac{(k+l+a)!\,b!}{(a+b+k+l+1)!}
\ee
Then after simple algebraic manipulations Eq.(\ref{jac1}) takes the form
\be
J = \dfrac{(a+1)_n(a+1)_m}{n!\,m!} \sum_{l} \dfrac{(-m)_l(a+b+m+1)_l\,a!\,b!\,2^{a+b+1}}{(a+b+l+1)_l\,l!} \\ \sum_k \dfrac{(-n)_k(a+b+n+1)_k(a+l+1)_k}{(a+b+l+2)_k(a+1)_k\,k!}
\ee
The last sum is a hypergeometric function $_3F_2(...|1)$.
To this function we apply famous Saalsch\"utz's identity
\be
\label{Sid}
{}_{3}F_2 \left(
\begin{array}{c}
a, b, -n  \\
c, 1+a+b-c-n
\end{array}\Big| 1
\right) = \dfrac{(c-a)_n(c-b)_n}{(c)_n(c-a-b)_n}
\ee
Therefore we get
\be
J = \dfrac{(a+1)_n(a+1)_m}{n!\,m!} \times \hspace{10cm} \\ \times \sum_{l} \dfrac{(-m)_l(a+b+m+1)_l\,a!\,b!\,2^{a+b+1}}{(a+b+l+1)_l\,l!} \dfrac{(-b-n)_n(-l)_n}{(a+1)_n(-a-b-n-l-1)_n}
\ee
Factors $(-m)_l$ and $(-l)_n$ give us non-vanishing condition
\be
\label{cond1}
n \leq l \leq m.
\ee
Then by simple algebraic manipulations we represent a sum over $l$ as a hypergeometric function $_2F_1(...|1)$ and apply identity (\ref{Sid}) once again:
\be
\label{jac2}
J =   \dfrac{ (a+1)_n(a+1)_m\,(a+b+m+1)_l\,a!\,b!\,2^{a+b+1} \,(-1)^n\,(-b-n)_n }{n!\,m!\,(a+1)_n(a+b+2n+1!\,(a+b+m)!\,(m-n)!} \times \\
\times \sum_{l} \dfrac{(n-m)_l(a+b+m+n+1)_l}{(a+b+2n+2)_l}  \hspace{5.4cm} \\
= \dfrac{ (a+1)_n(a+1)_m\,(a+b+m+1)_l\,a!\,b!\,2^{a+b+1} \,(-1)^n\,(-b-n)_n }{n!\,m!\,(a+1)_n(a+b+2n+1!\,(a+b+m)!\,(m-n)!} \times \\ \times \dfrac{(a+b+2n+1)!}{(n-m)!\,(a+b+n+m+1)!} \hspace{6cm}
\ee
It gives us second condition
\be
\label{cond2}
n\geq m.
\ee
Both conditions (\ref{cond1}) and (\ref{cond2}) imply
\be
n = m.
\ee
It finishes our proof of the orthogonality. In order to get r.h.s. of (\ref{jacort}) we put $n=m$ in formula (\ref{jac2}):
\be
J = \dfrac{(a+n)!\,(b+n)!\,2^{a+b+1}}{(a+b+n)!\,n!\,(a+b+2n+1)} \delta_{n,m}
\ee

\end{document}